\documentclass[journal]{IEEEtran}

\usepackage{xcolor}
\usepackage{color}

\usepackage{cite}
\usepackage[cmex10]{amsmath}
\usepackage{amssymb}
\usepackage{amsfonts}
\usepackage{algorithmic}
\usepackage{graphicx}
\usepackage{subfigure}
\usepackage{booktabs}
\usepackage{multirow}

\usepackage{bm}

\date{\today}

\begin{document}
	
	\title{Radar Accurate Localization of UAV Swarms Based on Range Super-Resolution Method}

	\author{Tianyuan Yang, Jibin Zheng, \textit{Member, IEEE}, Tao Su, and Hongwei Liu,\textit{ Member, IEEE}
	\thanks{T. Yang, J. Zheng, T. Su and H. Liu are with the National Laboratory of Radar Signal Processing, Xidian University, Xi’an, 710071, China. (Corresponding authors: Jibin Zheng and Tao Su. E-mail: jibin\_zheng@sina.cn;sutao\_xidian@163.com). }
	}
	\maketitle
	
	\begin{abstract}
		In radar accurate localization of unmanned aerial vehicle (UAV) swarms, the high density, similar motion parameters, small radar cross-section (RCS), strong noise and far range put forward high requirements on radar resolution and transmitting power. In this paper, by using advantages of the long-time integration (LTI) technique and gridless sparse method, we construct a super-resolution framework for radar accurate localization of UAV swarms without changing radar hardware and system parameters. Thereafter, based on this framework, a range super-resolution method is proposed to realize the radar accurate localization of UAV swarms. Mathematical analyses and numerical simulations are performed and demonstrate that, compared to the keystone transform (KT)-based LTI method, MUSIC-based method and reweighted atomic-norm minimization (RAM)-based method, the range super-resolution method is more robust and practical for radar accurate localization of UAV swarms under the noisy environment. Additionally, the real experiment with X-band radar is also conducted to verify the effectiveness of the range super-resolution method.
	\end{abstract}
	
	
	\IEEEpeerreviewmaketitle
	\section{Introduction}
	
	Because of the low cost, flexibility and convenience, the unmanned aerial vehicle (UAV) has attracted wide attention in both military and civilian areas, such as power line inspection, environmental monitoring, agriculture, battlefield surveillance, and so on \cite{wei2013operation,daponte2015metrology,bacco2016satellites,albani2017field}. With the development of communication and control technology, it is possible to replace a single UAV with UAV swarms which can bring many advantages, such as division of labor and cooperation and swarm intelligence\cite{wei2013operation}. Actually, it is more important for applications of UAV swarms in military. The battlefield environment is complex and changeable. The great flexibility of the UAV swarms can increase the probability of the successful attack\cite{wu2018modeling}.
	
	UAV swarms and the surrounding usually show characteristics of the high density, similar motion parameters, small radar cross-section (RCS), strong noise and far range which put forward high requirements on radar resolution and transmitting power for radar accurate localization of UAV swarms \cite{grossi2016new,xu2017focus}. Without changing radar hardware and system parameters, the long-time integration (LTI) technique is a good alteration to improve the Doppler resolution and decrease radar transmitting power. The range cell migration (RCM) and Doppler frequency migration are two main factors to influence the LTI \cite{xu2017focus}. In past decades, many excellent methods have been proposed to realize the LTI. Based on whether the phase information of echoes is used or not, the LTI methods can be categorized into the incoherent and coherent methods\cite{carlson1994search,carretero2009application,tandra2008snr,abatzoglou1998range,yu2012radon,zhu2007keystone,zheng2015radar,zheng2017parameterized}. Compared to the incoherent methods\cite{carlson1994search,carretero2009application}, the coherent methods use the phase information to guarantee a higher signal-to-noise ratio (SNR) improvement without the “SNR threshold”\cite{tandra2008snr}. The maximum likelihood estimation method and Radon-Fourier transform are both typical coherent methods\cite{abatzoglou1998range,yu2012radon}. In order to reduce the computational cost, the keystone transform (KT) is proposed and has been studied a lot to realize the effective coherent LTI\cite{zhu2007keystone,zheng2015radar,zheng2017parameterized}. Unfortunately, the LTI technique cannot improve the range and space resolutions. In addition, since the RCS scintillation may happen during a long illumination time, we cannot extend the integration time indefinitely to improve the Doppler resolution.
		
	It is necessary to develop the resolution improvement technique for radar accurate localization of UAV swarms without changing radar hardware and system parameters. The echo phase contains the range, Doppler and space information of UAV swarms. Therefore, the spectrum estimation technique may be used to improve radar resolution, such as the Capon’s beamforming and subspace methods like MUSIC \cite{schmidt1986multiple} and ESPRIT\cite{paulraj1986subspace}. However, this kind of methods needs sufficient samples and are sensitive to source correlations\cite{yang2018sparse,malioutov2005sparse}. In the last decade, due to the low sidelobe, high resolution and correlation insensitivity, the sparse technique has been greatly developed and can be categorized into on-grid sparse methods, off-grid sparse methods and gridless sparse methods\cite{yang2018sparse}. On-grid sparse methods assume the sparse signals are located on discrete grids\cite{malioutov2005sparse,donoho2009message,zhang2011iterative}. By contrast, off-grid sparse methods\cite{zhu2011sparsity_offgrid,fang2016super_offgrid,hu2013fast} take into account the deviation between the sparse signal and grid. However, they are still based on discrete gridding of the frequency domain. Gridless sparse methods are based on the atomic norm and can completely eliminate the grid mismatch\cite{tang2013compressed,yang2016exact,candes2014towards}. References \cite{yang2015enhancing} further proposed reweighted atomic-norm minimization (RAM) to break up the resolution limitation of gridless sparse methods. The current research on gridless sparse methods has shown the potential to improve the radar resolution under the noise-free environment without changing radar hardware and system parameters\cite{yang2018sparse}. Unfortunately, gridless sparse methods are sensitive to the degrees of freedom and low SNR.
	
	Analyses above indicate that, without changing radar hardware and system parameters, the LTI technique can increase the SNR and improve the Doppler resolution, and the gridless sparse method has the potential to improve the resolution. In addition, the LTI technique is easy to implement\cite{xu2017focus,zheng2015radar} and may provide the \textit{prior} knowledge of frequency recovery interval for the sparse gridless method. References\cite{mishra2015spectral} and\cite{yang2018frequency} have demonstrated that the \textit{prior} knowledge can yield a higher stability for the sparse gridless method in presence of noise. In this paper, by using advantages of the LTI technique and gridless sparse method, we construct a super-resolution framework for radar accurate localization of UAV swarms without changing radar hardware and system parameters. Thereafter, basing on this framework, we propose a range super-resolution method, which contains three steps and inherits advantages of the LTI technique and gridless sparse method. Mathematical analyses and numerical simulations are performed and demonstrate robustness and practicability of the proposed method. The real radar data is also sampled through an experiment with X-band radar and used to verify the effectiveness of the range super-resolution method.
	
	The remainder of this paper is organized as follows. Section \ref{System Model} gives the signal model and problem formulation. A novel super-resolution framework is presented in Section \ref{Proposed Fram}, where a range super-resolution method is also proposed. Section \ref{Computationally Efficient Implentation} presents some strategies for fast implementation of the proposed method. Section \ref{Numerical Simu} gives numerical simulations and real data processing result. Section \ref{Conclusion} includes the conclusion and future work.

	\section{Signal Model And Problem Formulation}  
	\label{System Model}
	
	Suppose radar transmits the linear frequency modulation continuous wave (LFMCW) signal. Note that the transmitting waveform is not just confined to the LFMCW signal waveform and can be other kinds of waveforms\cite{xu2017focus,zheng2015radar}. The LFMCW signal can be written as
	\begin{equation}\label{eq: transSig}
	s_{t}\left(\hat{t},t_m\right)=\operatorname{rect}\left( \frac{\hat{t}}{T}\right) \exp \left(j 2 \pi\left(f_{c} \hat{t}+\frac{\gamma}{2}\hat{t}^{2}\right)\right),
	\end{equation}
	where $\operatorname{rect}(\frac{\hat{t}}{T}) =\left\{\begin{array}{l}{1,|\hat{t}| \leq T / 2} \\ {0,|\hat{t}|>T / 2}\end{array}\right.$ is a rectangle window. $ \hat{t} $ is the fast time. $ t_m \in [-T_M/2,T_M/2] $ is the slow time, where $ T_M $ is the integration time. $ T, f_{c}$, and $\gamma$ denote the modulation period, carrier frequency, and frequency modulation rate, respectively. 
	\begin{figure}[h]
		\centering
		\includegraphics[height=6cm]{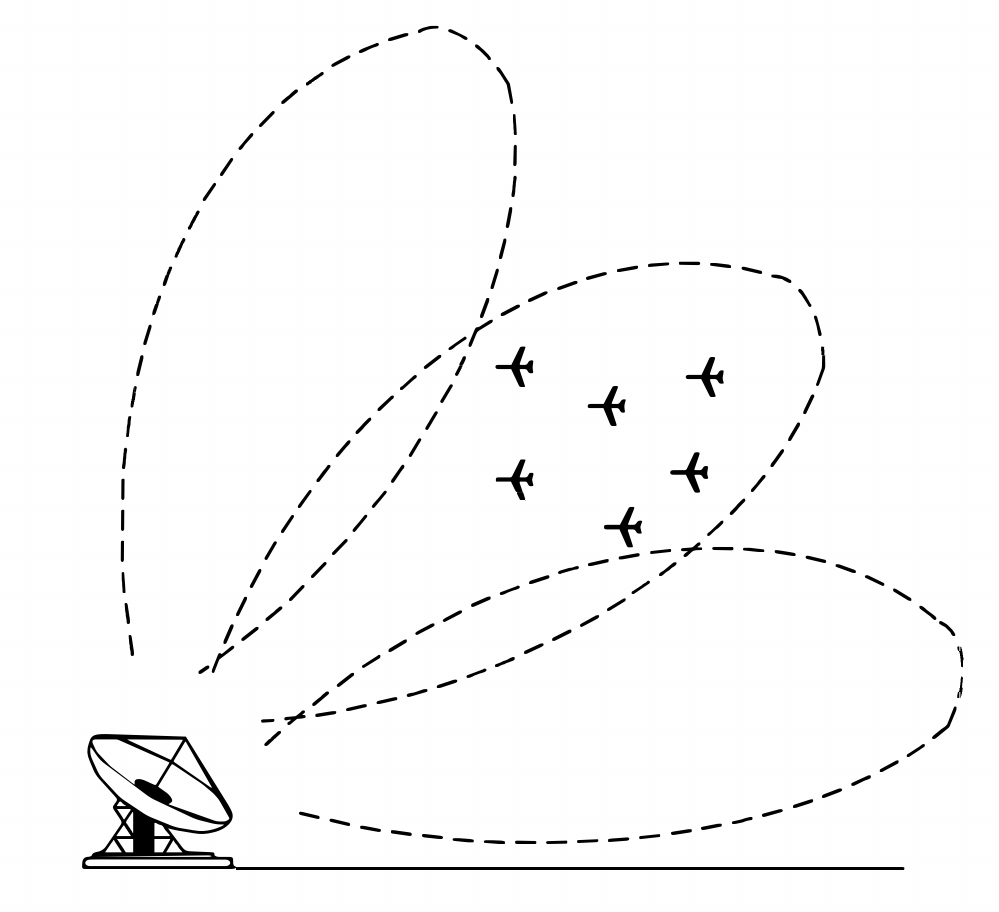}
		\caption{Illustration of radar localization of UAV swarms.}
		\label{fig:UAV}
	\end{figure}
	
	As shown in Fig. \ref{fig:UAV}, we assume there are $ K $  UAVs in one beam. For an  $ L $-element antenna array, the echo of $ K $ UAVs can be written as \cite{xu2017focus,zhang2018multitarget}
	\begin{equation}\label{eq: receiveSig}
	\begin{aligned}
	s_{r}&\left(\hat{t}, t_{m}, l\right)\\
	&=\sum_{k=1}^{K} A_{k}\operatorname{rect}\left( \frac{\hat{t}-\tau_{k}\left(t_{m}\right)}{T}\right) \\
	&\times\exp \left(j 2 \pi f_{c}\left(\hat{t}-\tau_{k}\left(t_{m}\right)\right)\right) \\  
	&\times \exp \left(j \pi \gamma\left(\hat{t}-\tau_{k}\left(t_{m}\right)\right)^{2}\right)\\
	& \times \exp \left(j \phi_{l}(\theta_{k})\right) +e_r\left(\hat{t}, t_{m}, l\right),
	\end{aligned}
	\end{equation}
	where
	\begin{equation}\label{eq:delay time}
	\begin{aligned}
	\tau_{k}\left(t_{m}\right)=\frac{2 R_{k}\left(t_{m}\right)}{c}
	\end{aligned}
	\end{equation}
	denote the  delay of the $ k $th UAV. 
	$ A_{ k},\theta_{k}, \phi_{l}(\theta_{k}) $ and $ c $ denote the amplitude, direction of the  $ k $th UAV, phase difference of the $ l $th antenna element from direction $ \theta_{k} $ and light velocity, respectively. $ e_r(\hat{t}, t_{m}, l) $ is the additive complex white Gaussian noise. 

	Considering the limited dynamical system and  the high density, we assume that the velocity and direction of each target are constant. The radial range $ R_{k}\left(t_{m}\right) $  between the  $ k $th UAV and radar at the slow time $ t_m $ can be expressed as
	\begin{equation} \label{R_k}
		R_{k}\left(t_{m}\right)=R_{k 0}+v_{k} t_{m},
	\end{equation}
	where $ R_{k 0},v_k $ and $\theta_{k}$ denote the initial radial range, velocity and direction of the $k$th UAV, respectively. Note that this proposed method may also be applied to the maneuvering UAV swarms by combining with the dechirp operation\cite{yu2012radon,zhu2007keystone,zheng2017parameterized}.
	
	Mixing the transmitted signal with the received signal and substituting $ R_{k}\left(t_{m}\right) $, we obtain the beat signal as 
	\begin{equation}\label{eq: beatFreq s}
	\begin{aligned}
	s_{beat}(\hat{t}, t_{m},l)&=\sum_{k=1}^{K} B_{ k} \operatorname{rect}\left( \frac{\hat{t}}{T}\right) \\
	&\times \exp (j 2 \pi f_c \tau_{k}(t_{m}) ) \exp \left(j 2 \pi\gamma \hat{t}\tau_{k}(t_{m}) \right)\\
	&\times \exp(-j\pi \gamma \tau^2_{k}(t_{m}))\\
	& \times \exp \left(j \phi_{l}(\theta_{k})\right)+e_{beat}\left(\hat{t}, t_{m}, l\right),
	\end{aligned}
	\end{equation}
	where $ B_k$ denotes the amplitude. $ e_{beat}(\hat{t}, t_{m}, l) $ is the additive complex white Gaussian noise.Since the delay $ \tau_{k}\left(t_{m}\right) $ is usually much smaller than $ T $, the $ \tau_{k}\left(t_{m}\right) $ in the rectangle window is omitted in (\ref{eq: beatFreq s})\cite{brooker2005understanding}.
	

	
	In the defense of UAV swarms, the number of UAVs, spatial distribution, formation and trajectories are crucial to determine the attack intention\cite{bacco2016satellites,wu2018modeling}. The accurate localization of UAV swarms is critical to obtain information above. Considering characteristics of UAV swarms and analyzing  (\ref{eq: beatFreq s}) carefully, we summarize two main challenges of radar accurate localization of UAV swarms and list them below.
	\begin{enumerate}
		\item \textit{\textbf{Limited radar resolution:}} The density of the UAV swarms is high and the motion parameters are similar. The radar range, Doppler and space resolutions subject to the bandwidth, integration time and antenna aperture\cite{xu2017focus}. For an equipped radar, range, Doppler and space resolutions may be limited to separate UAVs in a swarm.
		\item \textit{\textbf{Limited radar transmitting power:}} Compared with the large aircraft, the UAV's RCS is small (generally only 0.01$ m^2 $)\cite{pieraccini2017rcs}. In addition, the UAV is usually far away from the surveillance radar and the noise is strong. The low SNR of UAV swarms’ echoes puts forward a high request on radar transmitting power.
	\end{enumerate}

	The accurate defense of the UAV swarms puts forward high requirement on radar accurate localization of UAV swarms. If we resolve two challenges above through changing radar hardware and system parameters, the price will be too high. A signal processing method without changing radar hardware and system parameters is more preferred. In the following, we firstly construct a super-resolution framework, and then, a range super-resolution method is proposed for radar accurate localization of UAV swarms without changing radar hardware and system parameters.

	\section{Super-Resolution Framework and Range Super-Resolution Method}\label{Proposed Fram}
	\subsection{Super-resolution framework}\label{Super-resolution framework}
	The current research \cite{grossi2016new,xu2017focus,carlson1994search,carretero2009application,tandra2008snr,abatzoglou1998range,yu2012radon,zhu2007keystone,zheng2015radar,zheng2017parameterized,yang2016exact,candes2014towards,yang2015enhancing} indicates that, \textbf{i)} the LTI technique can reduce the requirement on radar transmitting power and increase the Doppler resolution, while it cannot change range and space resolutions when the radar system is fixed; \textbf{ii)} the gridless sparse method has the potential to improve radar resolution, while it is sensitive to the degrees of freedom and low SNR; \textbf{iii)} these inspire us that, if the LTI technique and the gridless sparse method can make use of mutual advantages and make up own shortages, two main challenges of radar accurate localization of UAV swarms can be addressed without changing radar hardware and system parameters. In this paper, under this guideline, we construct a super-resolution framework, as shown in Fig. 2. Its implementation is divided into three steps (we only list key processing of these three steps below):

	\begin{itemize}
		\item \textbf{Step 1:} Apply the LTI technique.
		\begin{itemize}
			\item \textbf{Purpose:} Detect the swarm and localize distinguishable UAVs.
			\item \textbf{Reason:} The illumination time is short during the searching stage and the SNR improvement of the LTI technique is limited. However, if we take the swarm as a target, its RCS will be larger than a single UAV and it will be easier to detect this swarm.
		\end{itemize}
		\item \textbf{Step 2:} Adjust radar and apply the LTI technique.
		\begin{itemize}
			\item \textbf{Purpose:} Localize distinguishable UAVs along the Doppler dimension, enhance the SNR and provide the \textit{prior} knowledge for the sparse gridless method.
			\item \textbf{Reason:} We can adjust radar to increase the illumination time and power after the swarm detection in \textbf{Step 1}. After applying the LTI technique, the \textit{prior} knowledge will be more accurate and improvements of Doppler resolution and SNR will be more significant than those of \textbf{Step 1}, which benefits the accurate localization along Doppler dimension and the sparse gridless method in \textbf{Step 3}. 
		\end{itemize}
		\item \textbf{Step 3:} Extract signal and apply the sparse gridless method.
		\begin{itemize}
			\item \textbf{Purpose:} Localize all potential UAVs along the range and space dimensions.
			\item \textbf{Reason:} We can transform the extracted signal into the superposition of single-frequency signals via the inverse Fourier transform along the range/space dimension. In addition, \textbf{Step 2} can provide \textit{prior} knowledge of frequency recovery interval and guarantee the high SNR. Therefore, the sparse gridless method can be used to localize all potential UAVs with a high robustness.
		\end{itemize}
	\end{itemize}

	Above is the constructed super-resolution framework for radar accurate localization of UAV swarms. It is worthwhile noting that this framework is not a simple combination of the LTI technique and sparse gridless method, but a deep fusion from aspects of the SNR enhancement, \textit{prior} knowledge of recovery frequencies, high Doppler resolution, more degrees of freedom, and so on. We can refer to the following subsection for more details about this deep fusion.
	\begin{figure}[h]
		\centering
		\includegraphics[height=6cm]{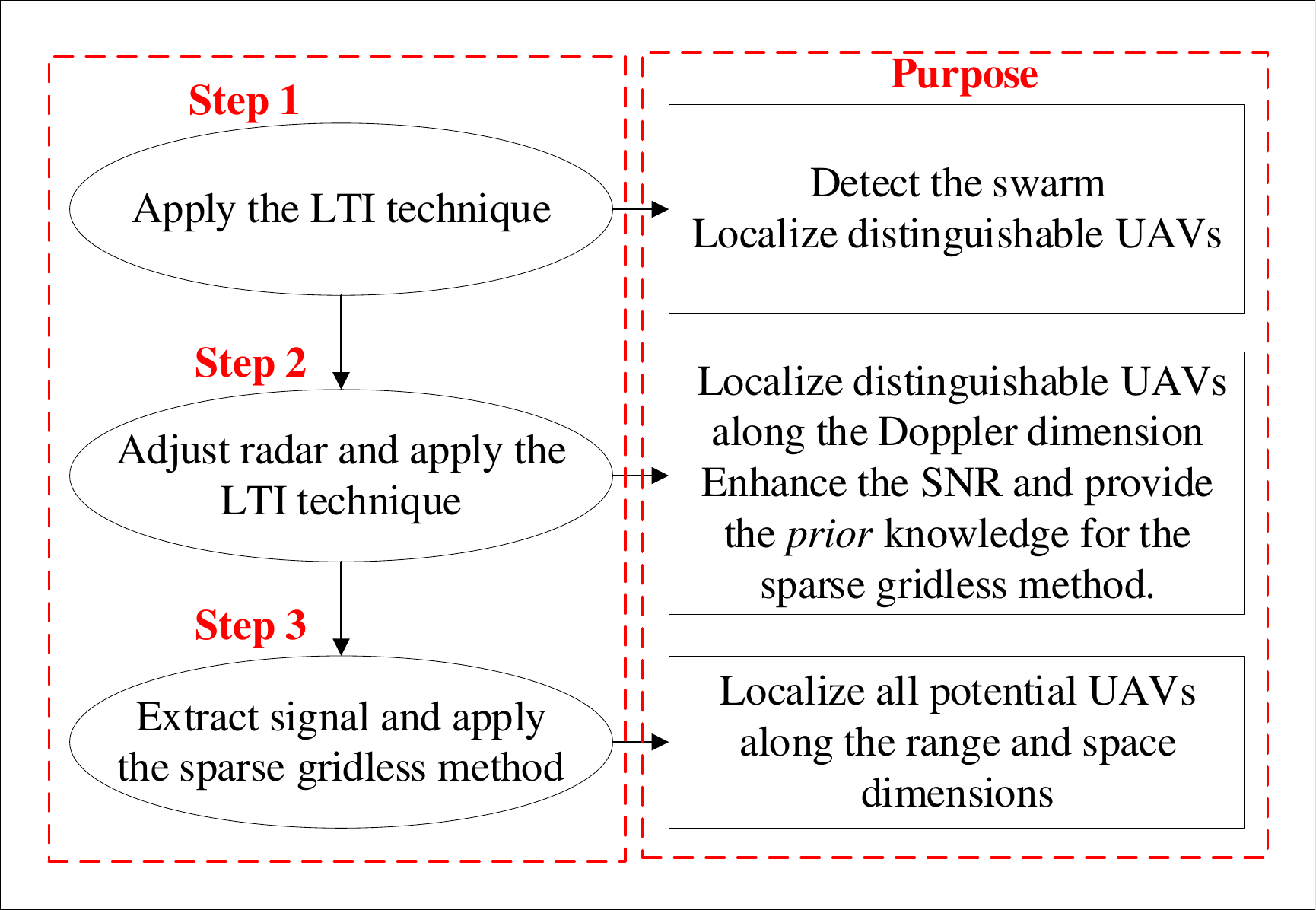}
		\caption{Structure of super-resolution framework.}
		\label{fig:3StepStruc}
	\end{figure}
	
	\subsection{Range super-resolution method}
	Basing on the framework constructed in the last subsection, we propose a range super-resolution method for radar accurate localization of UAV swarms in this subsection. Three steps of the proposed method and corresponding mathematical analyses are listed as follows. 
	
	\textbf{Step 1: Collect radar data, use the beamforming operation and KT-based LTI method to integrate the echo energy, detect distinguishable UAVs via the constant false alarm detection (CFAR), and then, estimate the velocity, range and angel.}
	
	Similar to the \textbf{Step 1} of the framework constructed in Subsection \ref{Super-resolution framework}, the key processing of this step is the KT-based LTI method and its main purpose is to find the swarm. Under the low echo SNR and short illumination time, this step makes full use of characteristics of the swarm and the KT-based LTI method maximizes the SNR improvement. Based on the sapling scheme of Claasen and Mecklenbrauker\cite{zheng2017parameterized} and combined with (\ref{eq:delay time}) and (\ref{R_k}), the discrete form of $ s_{beat}(\hat{t},t_m,l) $ can be expressed as
	\begin{equation}\label{eq: beat freq discrete}
	\begin{aligned}
	&S_{beat}(n, m,l)\\
	&=\sum_{k=1}^{K} C_{k} \operatorname{rect}\left( \frac{n\Delta\hat{t} }{T}\right) \times \exp \left(j 2 \pi\left(\gamma \frac{2 R_{k 0}}{c}\right) n\Delta\hat{t} \;\right) \\
	&\times \exp \left(j 2 \pi\left(\gamma \frac{2 v_{k }}{c}\right) mTn\Delta\hat{t} \;\right)\\
	&\times \exp \left(j 2 \pi\left(f_{c} \frac{2 v_{k }}{c}\right) mT\right)\\
	&\times \exp \left(j 2 \pi \left(f_{c}\frac{ld\sin(\theta_{k})}{c}\right)\right)+e_{beat}\left(n, m, l\right),
	\end{aligned}
	\end{equation}
	where $ C_{k}=B_{k}\exp(j2\pi f_c (2R_0/c) $. $ n=-N/2,-N/2+1,\ldots,N/2-1 $ and $ \Delta \hat{t} $ are the index of sampling and sampling interval of $\hat{t} $, respectively.
	$ m=-M/2,-M/2+1,\ldots,M/2-1$. $ N $ and $ M $ denote the number of samplings of fast time and slow time, respectively. Since $ v_k \ll c $ is very low, the exponential term $ \exp(-j\pi \gamma \tau^2_{k}(mT)) $  is ignored in (\ref{eq: beat freq discrete}). Here, we assume the antenna used here is a uniform linear array with two adjacent elements spaced by half the wavelength, i.e., $ d=\lambda/2=c/(2f_c) $.
	
	Then, we use the beamforming operation to integrate the energy in space dimension. Suppose $ G $  beam steering vectors are formed and the  $ g $th beam steering vector is written as 
	\begin{equation}\label{eq:steering vector}
	\bm{J}_g=[e^{-j\phi_1(\theta_{g})},e^{-j\phi_2(\theta_{g})},\ldots,e^{-j\phi_L(\theta_{g})}],
	\end{equation}
	where $ \phi_l(\theta_{g})=2\pi f_cld\sin(\theta_g)/c $. $ g=1,2,\ldots,G $ denotes the index of the beam steering vectors. Hence, the beamforming operation with $ \bm{J}_g $ is expressed as
	\begin{equation}\label{eq:BF}
	\begin{aligned}
	&S_{bf}(n, m,g)\\
	&=\sum_{l=1}^{L} S_{beat}(n, m,l) \times e^{-j\phi_l(\theta_{g})} \\
	&=\sum_{k=1}^{K} D_{k} \operatorname{rect}\left( \frac{n\Delta\hat{t} }{T}\right) \times \exp \left(j 2 \pi\left(\gamma \frac{2 R_{k 0}}{c}\right) n\Delta\hat{t} \;\right) \\
	&\times \exp \left(j 2 \pi\left(\gamma \frac{2 v_{k }}{c}\right) mTn\Delta\hat{t} \;\right)\\
	&\times \exp \left(j 2 \pi\left(f_c \frac{2 v_{k }}{c}\right) mT \;\right)\\
	&\times\operatorname{sinc}\left( \frac{L}{2}(\sin \theta_{k}-\sin \theta_{g})\right)+e_{bf}\left(n, m, g\right),
	\end{aligned}
	\end{equation}
	where  $ \operatorname{sinc}(x)=\sin(\pi x)/(\pi x) $.  $D_k= C_k\times L \exp(j\pi (\sin \theta_{k}-\sin \theta_{g})/2)$ and $ e_{bf}\left(n, m, g\right)$ denote the amplitude and noise after the beamforming operation, respectively. 
	
	The coupling between $ m $ and $ n $ in the second exponential term of (\ref{eq:BF}) induces the envelop offset. When the envelop offset exceeds the range resolution $ \Delta R =c/(2\gamma T) $, the RCM happens and influences the coherent LTI. The keystone transform (KT) can blindly compensate the RCM and can be speeded up via the chirp-z transform\cite{zhu2007keystone,zheng2015radar}. Here, we use the KT, i.e., $ \hat{m}=(1+\gamma n \Delta\hat{t}/f_c)m $ , to compensate the RCM.
	
	\begin{equation}\label{eq:KT}
	\begin{aligned}
	&S_{kt}(n, \hat{m},g)\\
	&=\sum_{k=1}^{K} D_{k} \operatorname{rect}\left( \frac{n\Delta\hat{t} }{T}\right)\times \exp \left(j 2 \pi\left(\gamma \frac{2 R_{k 0}}{c}\right) n\Delta\hat{t} \;\right) \\
	&\times \exp \left(j 2 \pi\left(f_c \frac{2 v_{k }}{c}\right) \hat{m}T \;\right)\\
	&\times\operatorname{sinc}\left( \frac{L}{2}(\sin \theta_{k}-\sin \theta_{g})\right)+e_{kt}\left(n, m, g\right),
	\end{aligned}
	\end{equation}
	where $ e_{kt}\left(n, m, g\right) $ denotes the noise after the KT operation.
	
	The coupling is eliminated in (\ref{eq:KT}) and we can use two-dimensional discrete Fourier transform to realize the energy accumulation as
	\begin{equation}\label{eq:KT FT}
	\begin{aligned}
	&S_{ktft}(n_{\hat{f}},n_{\hat{m}},g)\\
	&=\sum_{k=1}^{K} E_{k} \operatorname{sinc}\left( \frac{2\gamma N\Delta \hat{t}}{c}\left(R_{k0}-\frac{c}{2\gamma N \Delta \hat{t}} n_{\hat{f}}\right) \right) \\
	&\times\operatorname{sinc}\left( \frac{2MTc}{f_c}\left(v_k-\frac{c}{2MTf_c}n_{\hat{m}} \right) \right) \\
	&\times\operatorname{sinc}\left( \frac{L}{2}(\sin \theta_{k}-\sin \theta_{g})\right)+e_{ktft}\left(n_{\hat{f}},n_{\hat{m}},g\right),
	\end{aligned}
	\end{equation}
	where $ n_{\hat{f}}=-N/2,-N/2+1,\ldots,N/2-1 $ and $ n_{\hat{m}}=-M/2,-M/2+1,\ldots,M/2-1 $.  $ n_{\hat{f}} $ and $ n_{\hat{m}} $ denote samplings of frequency domains corresponding to range and velocity dimensions, respectively. $ E_k=D_k\times N\times M $ denotes the amplitude. $ e_{ktft}\left(n_{\hat{f}},n_{\hat{m}},g\right) $ denotes the noise after the KT operation.

	The signal energy in (\ref{eq:KT FT}) is coherently integrated at $ (2\gamma R_{k0} /c,2f_cv_k/c,\theta_{k} ) $. Now, the CFAR technique can be used to detect distinguishable UAVs (or find the swarm). Compare $ S_{ktft}(n_{\hat{f}},n_{\hat{m}},g) $ with the threshold value $ \eta_0 $ determined by the CFAR,
	\begin{equation}\label{eq:CFAR}
	\begin{aligned}
	\left| S_{ktft}(n_{\hat{f}},n_{\hat{m}},g)\right| \mathop{\gtrless}\limits_{H_0}^{H_1} \eta_0.
	\end{aligned}
	\end{equation}
	
	If $ \left| S_{ktft}(n_{\hat{f}},n_{\hat{m}},g)\right| $  is larger than $  \eta_0 $, there will be a target. Meanwhile, the range, velocity and angel can be directly estimated. On the contrary, if $\left| S_{ktft}(n_{\hat{f}},n_{\hat{m}},g)\right|$ is smaller than $  \eta_0 $, there will be no target.
	
	\textbf{Step 2: Let the radar illuminate the right angel , collect radar data, use the beamforming operation and estimated parameters as \textit{priori} knowledge to integrate the signal energy, detect distinguishable UAVs via the CFAR, and then, estimate radial range and velocity.}
	
	The \textbf{Step 1} can find the swarm and separate the distinguishable UAVs. However, due to the high density of UAV swarms and similar motion parameters, most UAVs cannot be separated in \textbf{Step 1}. Since the angel of the swarm has been estimated in \textbf{Step 1}, we can distribute more illumination time and power to the estimated angel (benefit the Doppler resolution and SNR after the LTI). Therefore, compared to \textbf{Step 1}, \textbf{Step 2} can significantly improve the Doppler and SNR. As long as UAVs have differences in radial velocities, they can be separated along the Doppler dimension in this step. In addition, this step also provides \textit{priori} knowledge which benefits the robustness of the sparse gridless method in \textbf{Step 3}.
	
	Let radar illuminate the estimated angle and collect the radar data. The beat signal can be expressed as
	\begin{equation}\label{eq: beat freq discrete2}
	\begin{aligned}
	&S'_{beat}(n, m',l)\\
	&=\sum_{k=1}^{K} F_{k} \operatorname{rect}\left( \frac{n\Delta\hat{t} }{T}\right) \times \exp \left(j 2 \pi\left(\gamma \frac{2 R'_{k 0}}{c}\right) n\Delta\hat{t} \;\right) \\
	&\times \exp \left(j 2 \pi\left(\gamma \frac{2 v'_{k }}{c}\right) m'Tn\Delta\hat{t} \;\right)\\
	&\times \exp \left(j 2 \pi\left(f_{c} \frac{2 v'_{k }}{c}\right) m'T\right)\\
	&\times \exp \left(j 2 \pi \left(f_{c}\frac{ld\sin(\theta_{k})}{c}\right)\right)+e'_{beat}\left(n, m', l\right),
	\end{aligned}
	\end{equation}
	where $ m'=-M'/2,-M'/2+1,\ldots,M'/2-1 $  and $M'$ denote the index and the number of slow time in this step, respectively. $ F_k $ and $ e'_{beat}\left(n, m', l\right) $ denote the amplitude and noise, respectively. Note that $ M'>M $ due to the longer illumination time. $ R'_{k0} $ and $v'_k $  denote the initial range and radial velocity of the  $ k $th target in the second round of the data collection, respectively.
	
	 We use the beamforming operation and KT operation with estimated parameters as \textit{priori} knowledge to integrate the signal energy of $ S'_{beat}(n, m',l) $ as
	 
 	\begin{equation}\label{eq:KT FT2}
	 \begin{aligned}
	 &S_{integer}(n_{\hat{f}},{n}'_{\hat{m}},g)\\
	 &=\sum_{k=1}^{K} H_{k}\operatorname{sinc}\left( \frac{2\gamma N\Delta \hat{t}}{c}\left(R'_{k0}-\frac{c}{2\gamma N \Delta \hat{t}} n_{\hat{f}}\right) \right) \\
	 &\times\operatorname{sinc}\left( \frac{2M'Tc}{f_c}\left(v'_k-\frac{c}{2M'Tf_c}n'_{\hat{m}} \right) \right) \\
	 &\times\operatorname{sinc}\left( \frac{L}{2}(\sin \theta_{k}-\sin \theta_{g})\right)+e_{integer}\left(n_{\hat{f}},n'_{\hat{m}},g\right),
	 \end{aligned}
	 \end{equation}
	 where $  n'_{\hat{m}}=-M/2,-M/2+1,\ldots,M/2-1 $. $ H_k $ and $ e_{integer}\left(n_{\hat{f}},n'_{\hat{m}},g\right)$ denotes the amplitude and noise, respectively.
	
	 The signal energy in (\ref{eq:KT FT2}) is coherently integrated at $ (2\gamma R'_{k0} /c,2f_cv'_k/c ,\theta_{k} )  $. Similar to the \textbf{Step 1}, we use the CFAR technique to detect distinguishable UAVs.
	 \begin{equation}\label{eq:CFAR2}
	 \begin{aligned}
	 \left| S_{integer}(n_{\hat{f}},{n}'_{\hat{m}},g)\right| \mathop{\gtrless}\limits_{H_0}^{H_1} \eta_1.
	 \end{aligned}
	 \end{equation}
	 
	 If  $ \left| S_{integer}(n_{\hat{f}},{n}'_{\hat{m}},g)\right| $ is larger than the threshold value $ \eta_1 $, there will be a target. Meanwhile, the radial range and velocity can be directly estimated. On the contrary, there will be no target.
	 
	 The long illumination time  guarantees a high Doppler resolution, while some UAVs may have the same radial velocity and still cannot be separated along the Doppler dimension in this step. Therefore, we need to use differences along the range and space dimensions to separate these UAVs in \textbf{Step 3}.

	 \textbf{Step 3: Extract the signal of the corresponding range- Doppler-space bin, apply two-dimensional discrete inverse Fourier transform, use the proposed frequency-selective RAM (FSRAM) to increase the range resolution to further separate potential UAVs, and then, estimate parameters of separated UAVs to complete the accurate localization.} 
	 
	 Two-dimensional discrete inverse Fourier transform along $ n_{\hat{f}} $ and $ g $ can transforms $S_{integer}(n_{\hat{f}},{n}'_{\hat{m}},g)  $ into superposition of single-frequency signals (multiple-measurement-vector case\cite{yang2018sparse}), while the advantages of the LTI technique are still maintained along the Doppler dimension. In addition, parameters estimated in \textbf{Step 2} provide frequency recovery interval which can serve as \textit{prior} knowledge to enhance the robustness of the sparse gridless method\cite{mishra2015spectral,yang2018frequency}. Under this guideline, we propose the FSRAM to further separate potential UAVs along the range dimension in this step.
	 
	 Basing on analyses above, we extract the signal of the $ p $th Doppler channel of $S_{integer}(n_{\hat{f}},{n}'_{\hat{m}},g)  $ [its energy is larger than in (\ref{eq:KT FT})] and apply the two-dimensional discrete inverse Fourier transform along $ \hat{f} $ and $ g $.
	 \begin{equation}\label{eq:signal picked}
	 \begin{aligned}
	 S_{p}(n,l)&=\sum_{q=1}^{Q}A_{cq}\exp\left( j2\pi\left( \gamma \frac{2R_{q0}}{c}n\Delta \hat{t}\right)  \right) \\&
	 \times\exp \left( j 2 \pi \left(f_{c}\frac{ld\sin(\theta_{q})}{c}\right)\right) +e_p(n,l),
	 \end{aligned}
	 \end{equation}
	 where  $ Q $, $ A_{cq} $ and $ e_p(n,l) $ denote the number of potential UAVs contained in the extracted channel, amplitude and noise, respectively. Note that the signal energy is still focused along the Doppler dimension in (\ref{eq:signal picked}), which indicates that the advantages of the LTI technique are still maintained.
	 
	 As shown in (\ref{eq:signal picked}), the signal takes the form of the superposition of single-frequency signals (multiple- measurement-vector case\cite{yang2018sparse}). In other word, the separation of potential UAVs along the range dimension can be converted to the frequency recovery. We rewrite (\ref{eq:signal picked}) into a form of matrix.

 	 \begin{equation}\label{eq:signal picked matrix}
	 \begin{aligned}
	 \bm{S}_p=\bm{Y}+\bm{E}=\bm{A}(f)\bm{X}+\bm{E},
	 \end{aligned}
	 \end{equation}
	 where $\bm{S}_p=\left[ \bm{s}_p(:,1),\cdots,\bm{s}_p(:,L)\right]\in \mathbb{C}^{N \times L}$ and $ s_p(:,l) \in \mathbb{C}^{N}$.
	 $\bm{A}(f)=[\bm{a}(f_1),\ldots,\bm{a}(f_Q)]\in \mathbb{C}^{N\times Q}$ and $ \bm{a}(f)=[1,e^{j2\pi f},\ldots,e^{j2\pi(N-1)f}]^T \in \mathbb{C}^N $, $ \bm{X}=\left[ \bm{x}(:,1),\cdots,\bm{x}(:,L)\right]\in \mathbb{C}^{Q \times L} $ and $ \bm{x}(:,l)\in \mathbb{C}^{Q}  $. $\bm{E}=\left[ \bm{e}_p(:,1),\cdots,\bm{e}_p(:,L)\right]\in \mathbb{C}^{N \times L}$ and $ \bm{e}_p(:,l) \in \mathbb{C}^{N}$.  $\mathbb{C},\bm{Y}\in \mathbb{C}^{N\times L}$ and $ [\cdot]^T $ is the set of complex numbers, noiseless signal and transpose operation, respectively.
	 
	 Since parameters estimated in \textbf{Step 2} provide frequency recovery interval which can serve as \textit{prior} knowledge to enhance the robustness of the sparse gridless method, we refer to ideas of the atomic norm minimization (ANM)\cite{tang2013compressed}, RAM\cite{yang2015enhancing} and frequency-selective ANM  (FSANM)\cite{yang2018frequency} to propose a novel sparse gridless method, known as FSRAM. 
	 
	 As shown in \cite{yang2018frequency}, the FS atomic $ \ell_0 $ norm can be defined as
	  \begin{equation}\label{eq:fs l0 norm}
	 \left\| \bm{Y} \right\|_{0}^{FS}=\inf_{f_w\in[f_L,f_H] ,\bm{\psi}_w}\left\lbrace \mathcal{K}:\bm{Y}=\sum_{w=1}^{\mathcal{K}} \bm{a}(f_{w})\bm{\psi}_w\right\rbrace,
	 \end{equation}
	 where $ \bm{\psi}_w \in \mathbb{C}^{1\times L} $ is the amplitude vector of the $ w $th sinusoid. $  [f_L,f_H] $ denotes a closed interval, i.e., the \textit{prior} knowledge of $f_w$. 
	 
	 Then, in the presence of bounded noise with $ \left\| \bm{E} \right\|_F\leq \eta  $ ,  we should solve the following problem for UAV signal
	 recovery.
  	\begin{equation}\label{eq:UAV signal recovery}
	 \begin{aligned}
	 \min_{\bm{Y}} &\left\| \bm{Y} \right\|_{0}^{FS},
	 \text { subject to } & \left\| \bm{S_p}-\bm{Y}\right\|_F\leq \eta,
	 \end{aligned}
	 \end{equation}
	 where $ \left\| \cdot \right\|_F$ denotes the Frobenius norms. It is noteworthy that $ \left\| \bm{Y} \right\|_{0}^{FS} $ directly exploits sparsity and breaks resolution limit. However, it is nonconvex and NP-hard to compute.
	 
	 Inspired by\cite{yang2015enhancing} and \cite{yang2018frequency}, we propose an algorithm termed as frequency-selective RAM (FSRAM) to solve the problem shown in (\ref{eq:UAV signal recovery}), which can be regarded as a locally convergent iterative algorithm.
	 \begin{equation}\label{eq:FS RAM}
	 \begin{aligned}
	 \min_{\bm{Z},\bm{Y},\bm{u}_i}& \frac{\sqrt{N}}{2} \operatorname{Tr}(\bm{W}_i\bm{T}(\bm{u}_i))+\frac{1}{2\sqrt{N}}\operatorname{Tr}(\bm{Z}), \\
	 \text { subject to } 
	 &{\left[ \begin{array}{cc}{\bm{Z}} & {\bm{Y}^H} \\ {\bm{Y}} & {\bm{T}(\bm{u}_i)}   \end{array}\right] \geq \bm{0}},
	 \left\| \bm{S_p}-\bm{Y}\right\|_F\leq \eta, \\
	 &\hat{\bm{T}} \geq \bm{0},\\
	 \end{aligned}
	 \end{equation}
	 where $ \bm{Z} $ is a free matrix variable. $ \operatorname{Tr}(\cdot) $ and $ (\cdot)^H $  denote the trace and conjugate
	 transpose of matrix, respectively. 	 
	 $ \bm{W}_i=(\bm{T}(\bm{u}_i)+\epsilon \bm{I})^{-1} $ and $ \bm{T}(\bm{u}_i) $  is the positive semidefinite Toeplitz matrix in the $ i $th iteration
	 \begin{equation}
	 	\bm{T}(\bm{u}_i)={\left[ 
	 	\begin{array}{cccc}
	 		{u_1}    & {u_2}& \cdots & {u_{N}} \\ 
	 		{\overline{u}_{2}} & {u_1}& \cdots & {u_{N-1}}\\
	 		{\vdots} & {\vdots}& \ddots & {\vdots}\\
	 		{\overline{u}_{N}} & {\overline{u}_{N-1}}& \cdots & {u_{1}}
	 	\end{array}\right]}\in \mathbb{C}^{N\times N}.
 	\end{equation}
 	And $ u_x $  is the  $ x $th element of  $ \bm{u}_i $. $ \overline{\cdot} $  denotes the conjugate operation. $ \epsilon>0 $ is a regularization parameter to control the sparse metric to approach $ \ell_{0} $  norm $ (\epsilon \rightarrow 0 ) $ or $ \ell_{1} $  norm (large $ \epsilon $)\cite{yang2015enhancing}.  	
 	$ \hat{\bm{T}} $ is a matrix whose elements are made up of weighted elements of $ \bm{T}(\bm{u}_j)$.
 	
	 \begin{equation}\label{eq:Tg}
	 \begin{aligned}
	 \hat{\bm{T}}=&h_1\times{\left[ 
	 	\begin{array}{cccc}
	 	{u_2}    & {u_3}& \cdots & {u_{N}} \\ 
	 	{\overline{u}_{1}} & {u_2}& \cdots & {u_{N-1}}\\
	 	{\vdots} & {\vdots}& \ddots & {\vdots}\\
	 	{\overline{u}_{N-2}} & {\overline{u}_{N-3}}& \cdots & {u_{2}}
	 	\end{array}\right]}\\
 	+&h_2\times{\left[ 
 		\begin{array}{cccc}
 		{u_1}    & {u_2}& \cdots & {u_{N-1}} \\ 
 		{\overline{u}_{2}} & {u_1}& \cdots & {u_{N-2}}\\
 		{\vdots} & {\vdots}& \ddots & {\vdots}\\
 		{\overline{u}_{N-1}} & {\overline{u}_{N-2}}& \cdots & {u_{1}}
 		\end{array}\right]}\\
 	+&\overline{h}_1\times{\left[ 
 		\begin{array}{cccc}
 		{\overline{u}_2}    & {u_1}& \cdots & {u_{N-2}} \\ 
 		{\overline{u}_{3}} & {\overline{u}_2}& \cdots & {u_{N-3}}\\
 		{\vdots} & {\vdots}& \ddots & {\vdots}\\
 		{\overline{u}_{N}} & {\overline{u}_{N-1}}& \cdots & {\overline{u}_2}
 		\end{array}\right]},
	 \end{aligned}
	 \end{equation}
	 where $ h_1=e^{j\pi (f_L+f_H)}\operatorname{sign}(f_H-f_L) $  and $ h_2=-2\cos(\pi (f_H-f_L))\operatorname{sign}(f_H-f_L) $  denote the weight coefficients. $ \operatorname{sign}(\cdot) $ denotes the sign function. We can refer to\cite{mishra2015spectral} and\cite{yang2018frequency} for more details about (\ref{eq:Tg}). Here, the \textit{prior} knowledge can be written as
	 \begin{equation}\label{eq:prior}
	  \begin{aligned}
	  &{f_L=\frac{2\Delta \hat{t} \gamma (R_{est\_min}-\Delta R)}{c}}, \\
	  &{f_H=\frac{2\Delta \hat{t} \gamma (R_{est\_max}+\Delta R)}{c}},
	   \end{aligned}
	 \end{equation}
      where $ \Delta R=c/(2\gamma T) $ denotes the range resolution. $ R_{est\_min} $ and $ R_{est\_max} $ denote the range resolution, the nearest and furthest   range of the UAV(s) obtained in \textbf{Step 2}, respectively.
	 
	 By making the regularization parameter $ \epsilon $ approach to 0 gradually, the sparse metric approaches the atomic $ \ell_{0} $ norm with the highest frequency resolution\cite{yang2015enhancing}. The improvement of frequency resolution corresponds to the improvement of range resolution. Once we obtain the optimizer $ \bm{T}(\bm{u}^*) $, the frequency and accurate range can be retrieved using the Vandermonde decomposition, i.e., 
 	 \begin{equation}\label{eq:VD}
	 \left\{
	 \begin{aligned}
	  \bm{T}(\bm{u}^*)&=\sum_{q=1}^{Q}\sigma_q\bm{a}(f_q)\bm{a}^H(f_q),\\
	  R_{q0}&=\frac{cf_{q}}{2\gamma\Delta t },
	 \end{aligned}
	 \right.
	 \end{equation}
	 where $ \sigma_q>0 $. 
	 Note that, compared to the RAM proposed in \cite{yang2015enhancing}, the proposed FSRAM considers the \textit{priori} knowledge obtained in \textbf{Step 2} which corresponds to third constraint in (\ref{eq:FS RAM}). Referring to the ANM and FSANM, we know that this can yield a higher stability in presence of noise\cite{mishra2015spectral}. 
	 
\begin{figure}[h]
	\centering
	\includegraphics[height=5.5cm]{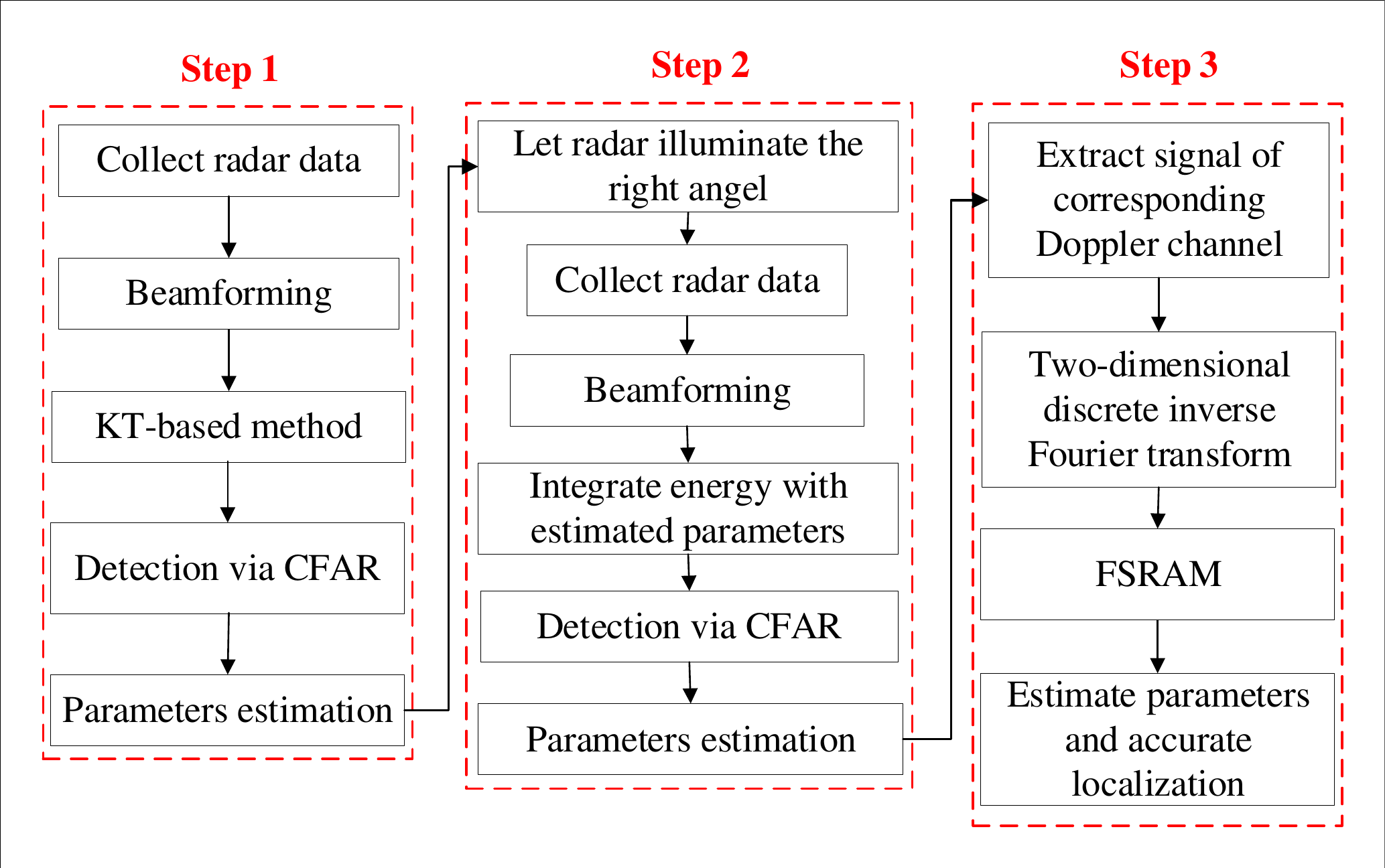}
	\caption{Structure of the proposed range super-resolution method.}
	\label{fig3}
	\end{figure}

	 \textbf{Steps 1, 2} and \textbf{3} constitute the proposed range super-resolution method. In summary, \textbf{Steps 1} aims to find the swarm, \textbf{Steps 2} aims to enhance the echoes’ SNR and separate adjacent UAVs in the Doppler dimension, and \textbf{Steps 3} aims to complete the separation of adjacent UAVs in the range dimension. If we still want to further separate UAVs in the space dimension, we can repeat \textbf{Steps 3} to recover the frequency corresponding to the accurate angel. Actually, most UAVs can be separated along the Doppler and space domains in realistic applications \cite{bacco2016satellites,albani2017field,wu2018modeling,grossi2016new,xu2017focus,yu2012radon,zhu2007keystone,zheng2015radar}. To better illustrate the proposed range super-resolution method, Fig.\ref{fig3} gives its structure and a numerical experiment is given below also.

\begin{figure*}[h]
	\subfigure[]{
		\label{fig4a} 
		\includegraphics[width=55mm]{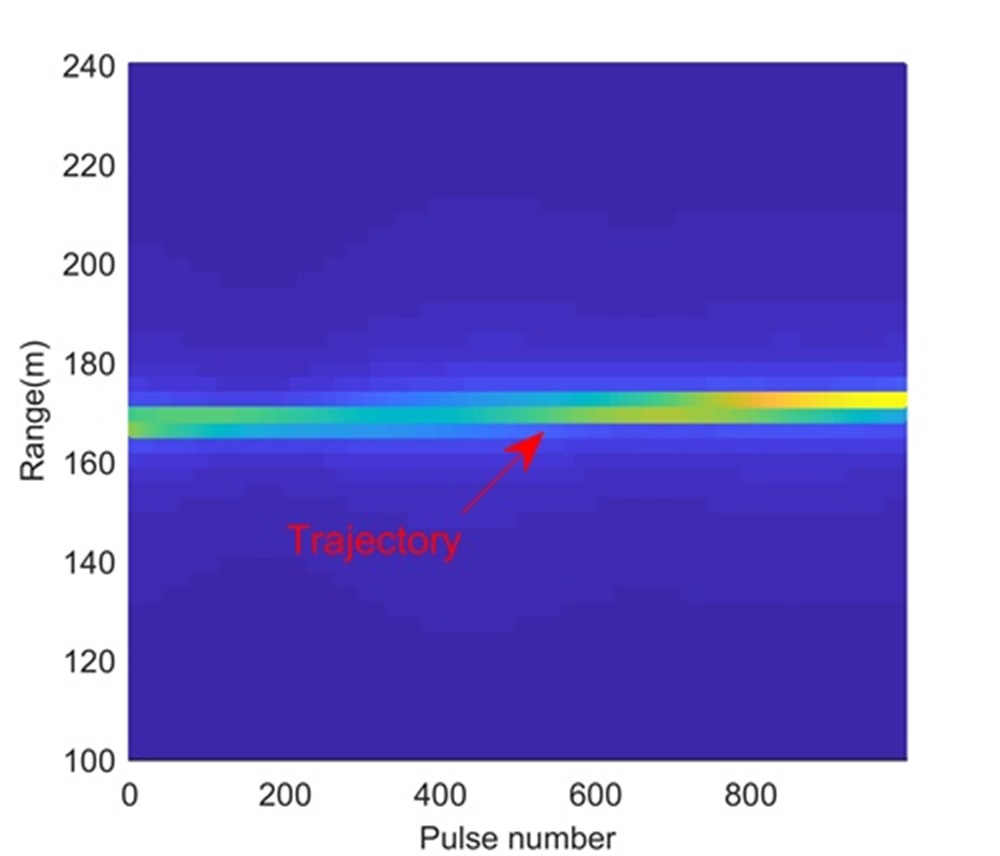}}
	\subfigure[]{
		\label{fig4b}
		\includegraphics[width=55mm]{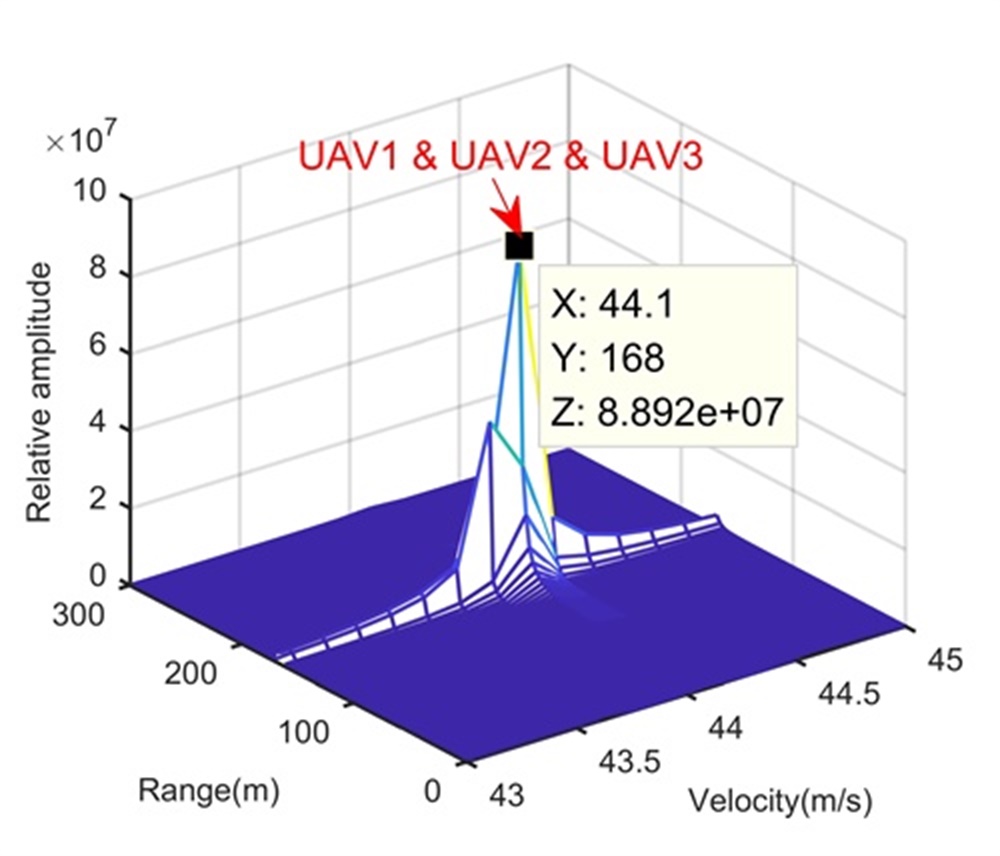}}
	\subfigure[]{
		\label{fig4c}
		\includegraphics[width=55mm]{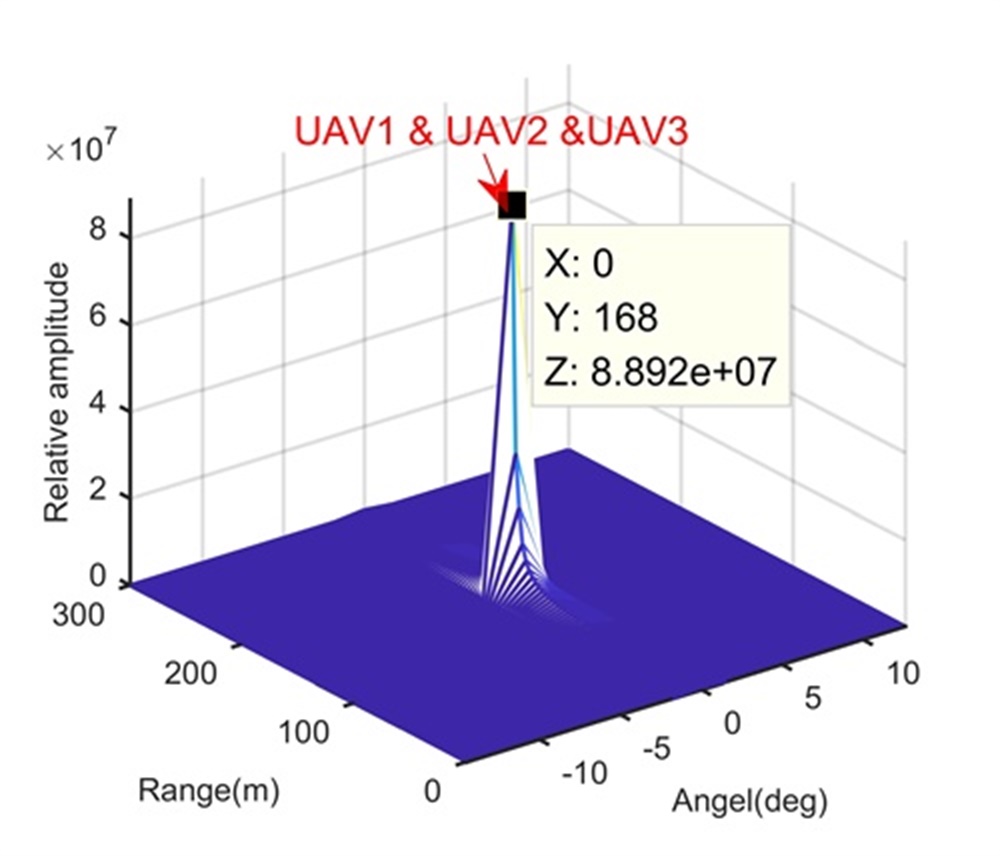}}
	
	\subfigure[]{
		\label{fig4d}
		\includegraphics[width=55mm]{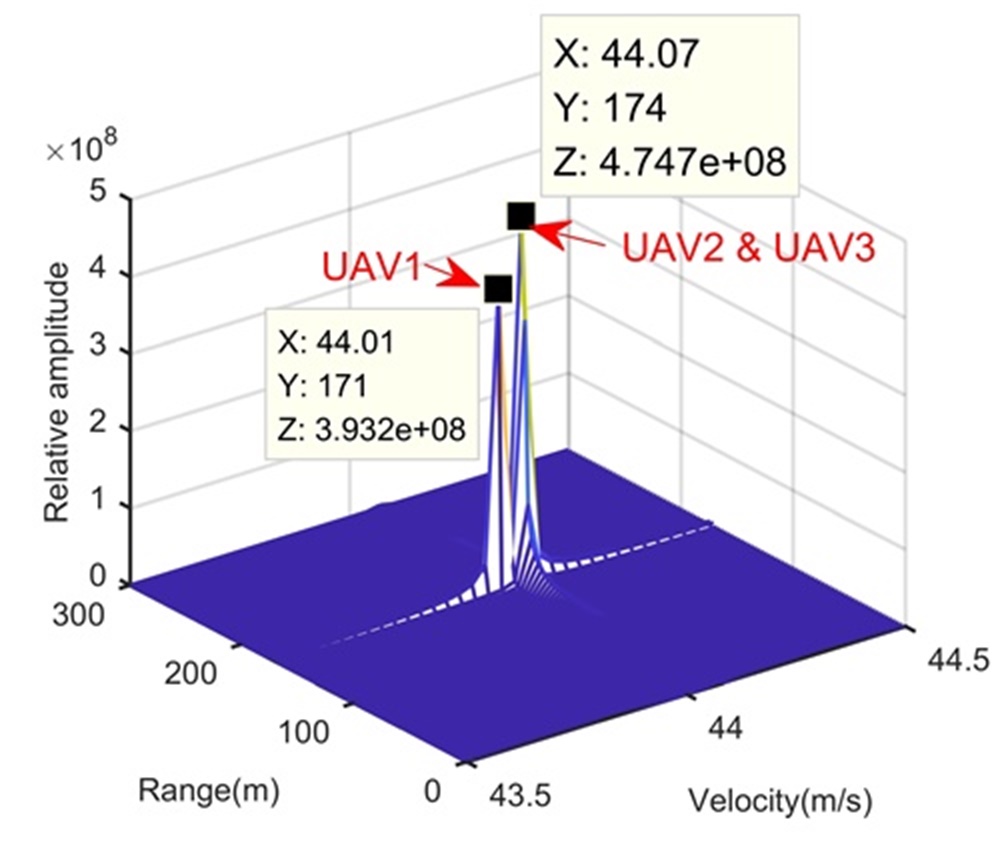}}
	\subfigure[]{
		\label{fig4e}
		\includegraphics[width=55mm]{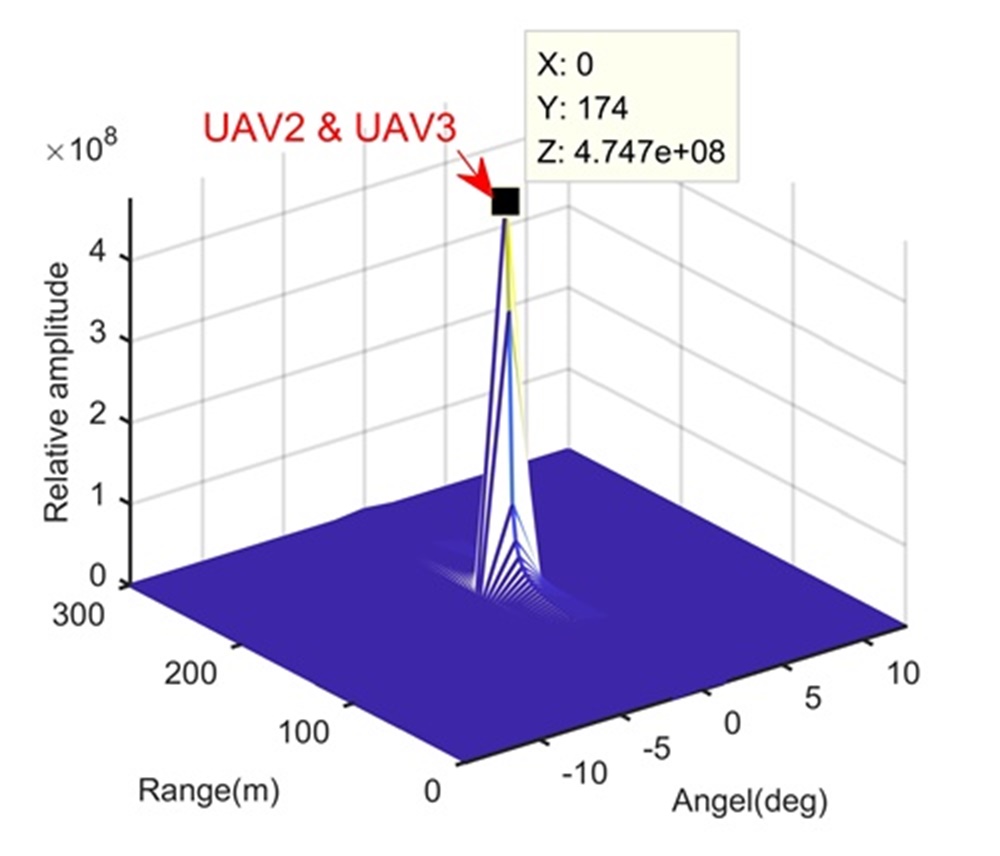}}
	\subfigure[]{
		\label{fig4f}
		\includegraphics[width=55mm]{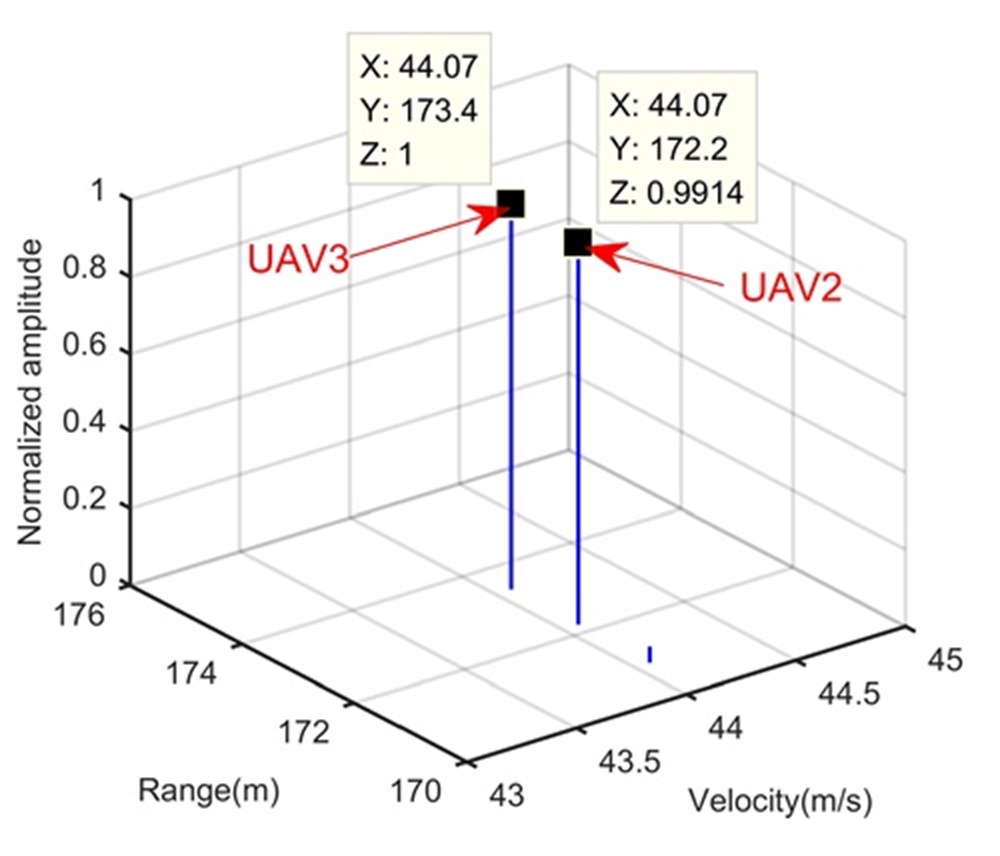}}
	\caption{Illustrate how the proposed range super-resolution method works. (a) Trajectory in the range-slow time domain after the beamforming. (b) Integration result in the range-Doppler domain after  \textbf{Step 1}. (c) Integration result in the range-space domain after  \textbf{Step 1}. (d) Integration result in the range-Doppler domain after  \textbf{Step 2}. (e) Integration result in the range-space domain after  \textbf{Step 2}. (f) Processing result after  \textbf{Step 3}.}
	\label{fig:Exp1 Result} 
\end{figure*}

	\textit{\textbf{Experiment 1:}} We consider three UAVs with constant velocities, same echo amplitude and same direction. We set radar illumination times 0.1s and 0.5s for \textbf{Steps 1} and \textbf{2}, respectively. The radar and target parameters are listed in Table \ref{tab:Radar para Exp1} and \ref{tab:Target para Exp1}, respectively. Fig. \ref{fig:Exp1 Result} shows simulation results.
			\begin{table}[h]
		\caption{Radar Parameters}
		\label{tab:Radar para Exp1}
		\begin{tabular}{ccclc}
			\hline
			\multicolumn{5}{c}{\textbf{Radar parameter}}                                                                                  \\ \hline
			\multicolumn{1}{c|}{Carrier frequency} & \multicolumn{1}{c|}{10GHz} & \multicolumn{2}{c|}{Sampling frequency}         & 50MHz \\ \hline
			\multicolumn{1}{c|}{Bandwidth}         & \multicolumn{1}{c|}{50MHz} & \multicolumn{2}{c|}{Number of antenna elements} & 16    \\ \hline
			\multicolumn{1}{c|}{Chirp duration}    & \multicolumn{1}{c|}{100us} & \multicolumn{2}{c|}{Range resolution}           & 3m    \\ \hline
		\end{tabular}
	\end{table}
	\begin{table}[h]
		\caption{Target Parameters}
		\begin{tabular}{c|c|c|c|c}
			\hline
			\multirow{2}{*}{} & \multicolumn{2}{c|}{\textbf{Step1}} & \multicolumn{2}{c}{\textbf{Step2}} \\ \cline{2-5} 
			& Range(m)       & Velocity(m/s)      & Range(m)       & Velocity(m/s)      \\ \hline
			UAV1              & 165.00         & 44.01              & 171.00         & 44.01              \\ \hline
			UAV2              & 166.20         & 44.07              & 172.20         & 44.07              \\ \hline
			UAV3              & 167.40         & 44.07              & 173.40         & 44.07              \\ \hline
		\end{tabular}\label{tab:Target para Exp1}
	\end{table}

	Fig. \ref{fig4a} shows the trajectory in the range-slow time domain after the beamforming. During the illumination, the RCM happens and may influence the LTI. We perform the KT-based LTI method to integrate the signal energy, and Figs.\ref{fig4b} and \ref{fig4c} show integration results in the range-Doppler domain and range-space domain, respectively. Because of the limited range, Doppler and space resolutions, although the processing in \textbf{Step 1} integrates the signal energy, it cannot separate these three UAVs. We apply the CFAR technique to the integration result of \textbf{Step 1} to detect the target (or find the swarm) and estimate parameters simultaneously for \textbf{Step 2}. After the processing in \textbf{Step 2}, Figs. \ref{fig4d} and \ref{fig4e} show integration results in the range-Doppler domain and range-space domain, respectively. In this step, due to the longer integration time, the Doppler resolution is increased and UAV1 is separated from UAV2 and UAV3. However, UAV2 and UAV3 are still not be separated. According to the proposed range super-resolution method, we extract the signal of the Doppler channel containing UAV2 and UAV3, and apply  \textbf{Step 3}. The processing result is given in Fig. \ref{fig4f} where UAV2 and UAV3 are successfully separated along the range dimension. Now, with parameters estimated in Figs. \ref{fig4d}, \ref{fig4e} and \ref{fig4f}, we complete radar accurate localization of the UAV swarms.

	\section{Computationally Efficient Implementation}\label{Computationally Efficient Implentation}
	The key processing of the proposed range super-resolution method is the KT-based LTI method and the FSRAM. In this section, we discuss computationally efficient implementations of the key processing.
	\subsection{Implementation of KT-based LTI method}
	Actually, operations in (\ref{eq:KT}) and (\ref{eq:KT FT}) determine the efficiency of the KT-based LTI method. Referring to \cite{zhu2007keystone,zheng2015radar}, we know that the KT operation in (\ref{eq:KT}) is only for the intuitive illustration of the function of the KT. The KT in (\ref{eq:KT}) and the integration along $ m $ (\ref{eq:KT FT}) can be done together via the scaled Fourier transform \cite{zhu2007keystone,zheng2015radar} which can be written as
	\begin{equation}
	\begin{aligned}\label{eq:kt1ft}
	&S_{kt\_1\_ft}(n,n_{\hat{m}},g)\\
	&=	\sum_{m}S_{bf}(n,m,g)\exp\left(-j2\pi \frac{n_{\hat{m}}}{MT} \left( 1+\frac{\gamma n\Delta \hat{t}}{f_c}\right)mT  \right) 
	\end{aligned}
	\end{equation}
	
	Basing on the characteristic of the formula (\ref{eq:kt1ft}), we can rewrite it as
	\begin{equation}\label{eq:kt1ft 2}
	\begin{aligned}
	&S_{kt\_1\_ft}(n,n_{\hat{m}},g)\\
	&=	\exp\left( -j\pi  \left( 1+\frac{\gamma n\Delta \hat{t}}{f_c}\right)\left( \frac{n_{\hat{m}}}{MT}\right) ^2 \right) \sum_{m}S_{bf}(n,m,g)\\
	&\times \exp\left(-j\pi \left( 1+\frac{\gamma n\Delta \hat{t}}{f_c}\right)(mT)^2  \right) \\
	&\times \exp\left(j\pi \left( 1+\frac{\gamma n\Delta \hat{t}}{f_c}\right)\left( \frac{n_{\hat{m}}}{MT}-mT\right) ^2  \right) 
	\end{aligned}
	\end{equation}
	
	The summation in (\ref{eq:kt1ft 2}) can be regarded as a convolution. So we can implement it with the FFT.
	\begin{equation}
	\begin{aligned}
	&S_{kt\_1\_ft}(n,n_{\hat{m}},g)\\
	&=\hat{h}(n,n_{\hat{m}},g)\times \operatorname{FFT}_m\left\lbrace \exp\left(j\pi  \left( 1+\frac{\gamma n\Delta \hat{t}}{f_c}\right)(mT)^2  \right)\right\rbrace\\
	& \times\operatorname{FFT}_m\left\lbrace S_{bf}(n,m,g)\exp\left(-j\pi \left( 1+\frac{\gamma n\Delta \hat{t}}{f_c}\right)(mT)^2  \right)\right\rbrace,\\
	\end{aligned}
	\end{equation}
	where $ \hat{h}(n,n_{\hat{m}},g)=\exp\left(-j\pi \left( {n_{\hat{m}}}/{MT}\right) ^{2} \left( 1+{\gamma n\Delta \hat{t}}/{f_c}\right) \right) $.
	
	Subsequently, the FFT can be used to integrate the energy along  $ n $, which can be presented as
	\begin{equation}
	\begin{aligned}
	S_{kt\_ft}(n_{\hat{f}},n_{\hat{m}},g)=\operatorname{FFT}_n\left\lbrace S_{kt\_1\_ft}(n,n_{\hat{m}},g)\right\rbrace,\\
	\end{aligned}
	\end{equation}
	
	Above is the computationally efficient implementations of (\ref{eq:KT}) and (\ref{eq:KT FT}) which only need FFT operations.
	
	\subsection{Implementation of FSRAM}
	As described in (\ref{eq:FS RAM}), the FSRAM can be reformulated as an exact semidefinite program. Since the dual problem can be solved more efficient than the primal problem via the standard SDP solver SDPT3\cite{toh1999sdpt3}, we can solve the dual problem of (\ref{eq:FS RAM}), and then, the solutions of the primal problem can be given for free. Meanwhile, some scholars proposed a first-order algorithm for the SDP based on ADMM\cite{boyd2011distributed}. If a more relaxed convergence criterion is adopted, the ADMM can be further accelerated. However, when it is close to the optimal value, the ADMM converges slowly\cite{yang2015enhancing}. Therefore, for the ADMM, there is a trade-off between accuracy and the number of iterations. We can refer to\cite{boyd2011distributed} for more details.
	
	\section{Numerical Simulations and Real Data Processing Results}\label{Numerical Simu}
	In this section, we compare the proposed range super-resolution method with the LTI method, MUSIC-based method and RAM-based method from two aspects, radar accurate localization of UAV swarms [\textit{\textbf{Experiment 2}} (noise-free environment) and \textit{\textbf{Experiment 3}} (noisy environment)] and success rate of adjacent UAVs’ separation [\textit{\textbf{Experiment 4}} (fixed SNR) and \textit{\textbf{Experiment 5}} (different SNRs)]. Note that the MUSIC-based method and RAM-based method only use one modulation period of $ S'_{beat}(n,m',l) $. This is because the coupling between $ n $ and $ m' $ lets $ S'_{beat}(n,m',l) $  not take the form of the superposition of single-frequency signals along $ n $. In addition, the real data is also used to verify the effectiveness of the proposed method.
	\subsection{Radar accurate localization of UAV swarms}
	In this subsection, the LTI method, MUSIC-based method, RAM-based method and proposed range super-resolution method are used for radar UAV swarms detection. Here, we use two experiments to illustrate accurate localization of UAV swarms for these four methods. \textit{\textbf{Experiment 2}} (noise-free environment) and \textit{\textbf{Experiment 3}} (noisy environment).
	\begin{table}[h]
	\caption{Target Parameter}
	\begin{tabular}{c|c|c|c|c}
		\hline
		\multirow{2}{*}{} & \multicolumn{2}{c|}{\textbf{Step1}} & \multicolumn{2}{c}{\textbf{Step2}} \\ \cline{2-5} 
		& Range(m)       & Velocity(m/s)      & Range(m)       & Velocity(m/s)      \\ \hline
		UAV4              & 162.00         & 44.01              & 168.00         & 44.01              \\ \hline
		UAV5              & 162.00         & 44.13              & 168.00         & 44.13              \\ \hline
		UAV6              & 163.20         & 44.13              & 169.20         & 44.13              \\ \hline
		UAV7              & 164.40         & 44.13              & 170.40         & 44.13              \\ \hline
	\end{tabular}\label{tab:Target para Exp2}
	\end{table}
	
\begin{figure*}[h]
		\subfigure[]{
			\label{fig5a} 
			\includegraphics[width=55mm]{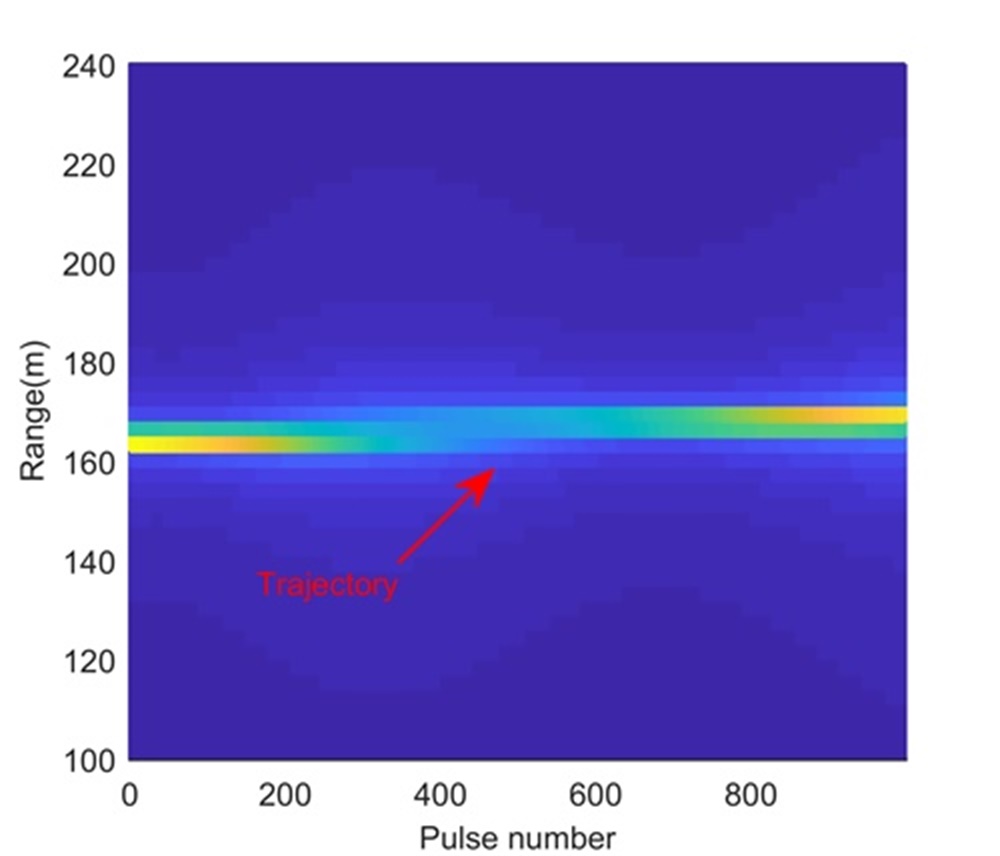}}
		\subfigure[]{
			\label{fig5b}
			\includegraphics[width=55mm]{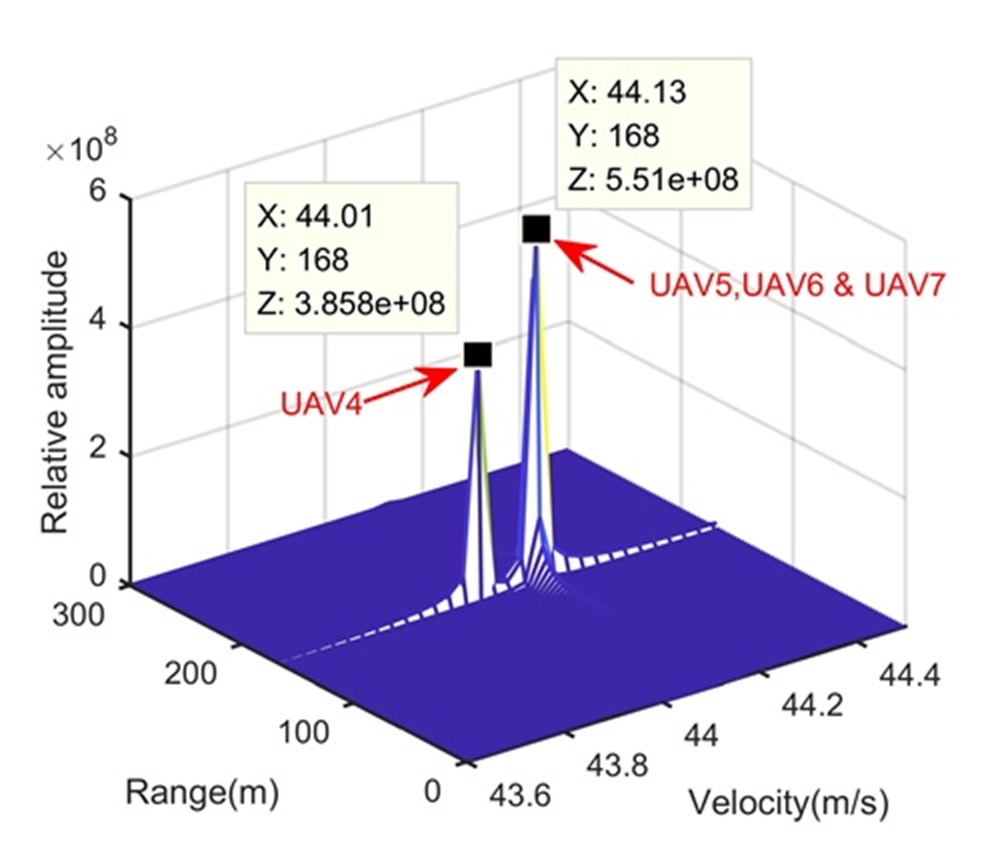}}
		\subfigure[]{
			\label{fig5c}
			\includegraphics[width=55mm]{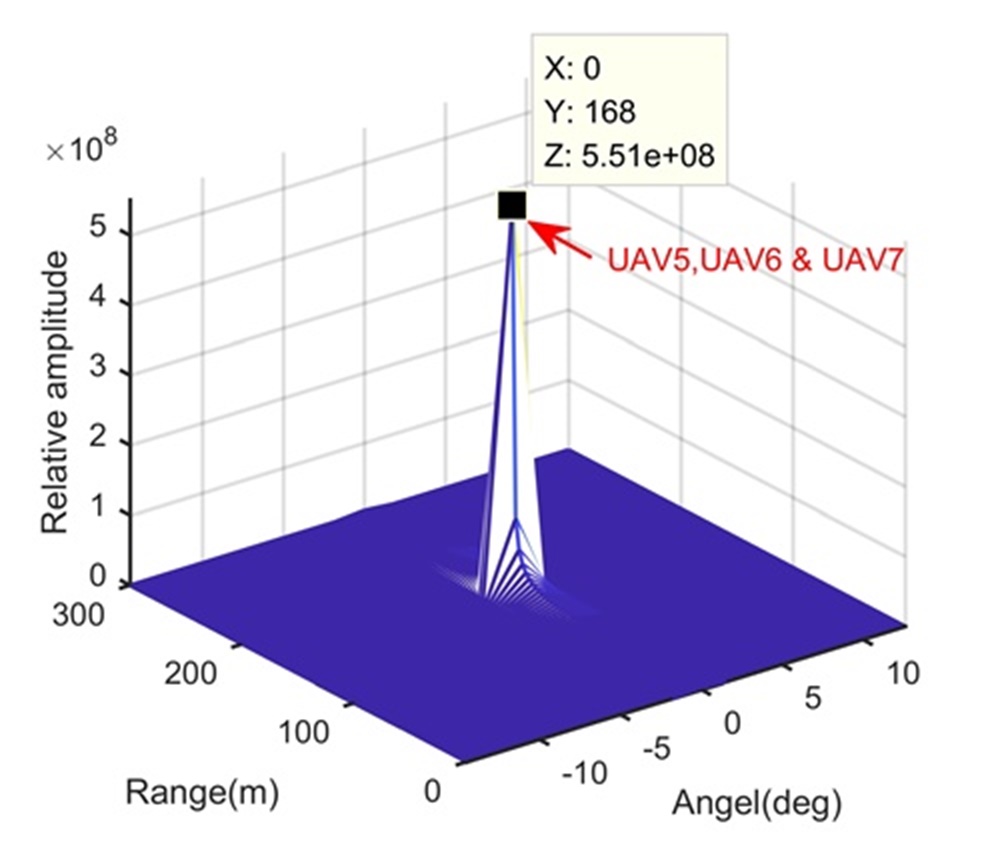}}
		
		\subfigure[]{
			\label{fig5d}
			\includegraphics[width=55mm]{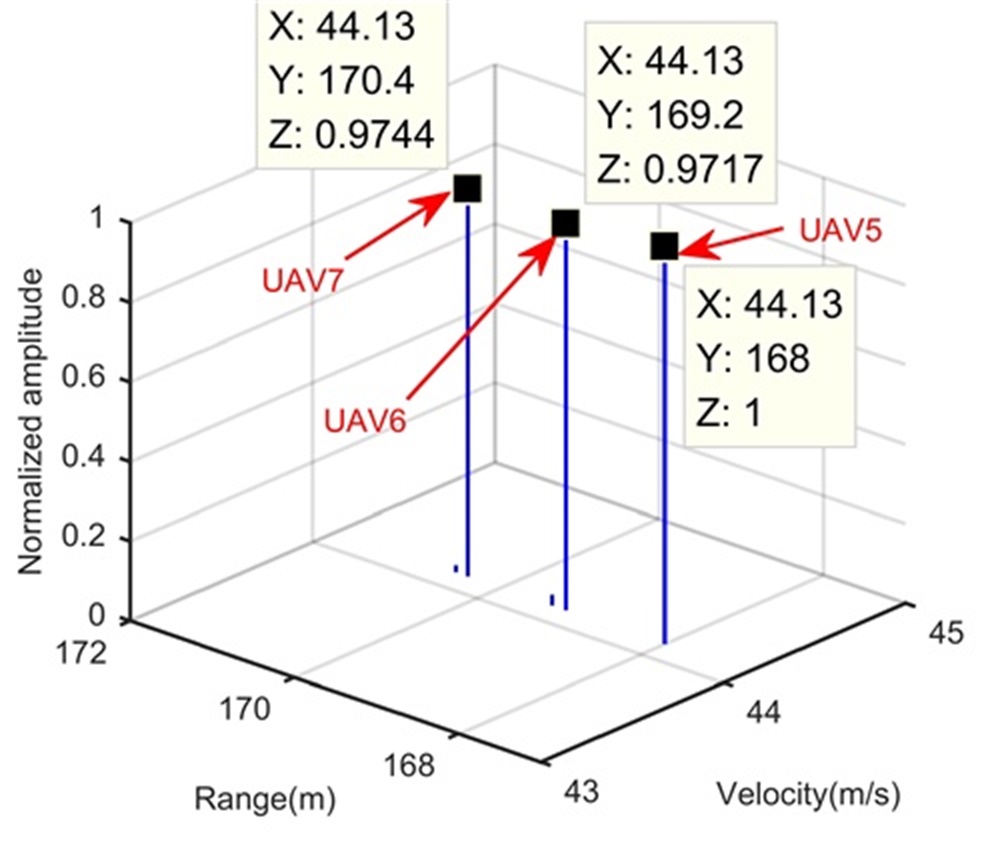}}
		\subfigure[]{
			\label{fig5e}
			\includegraphics[width=55mm]{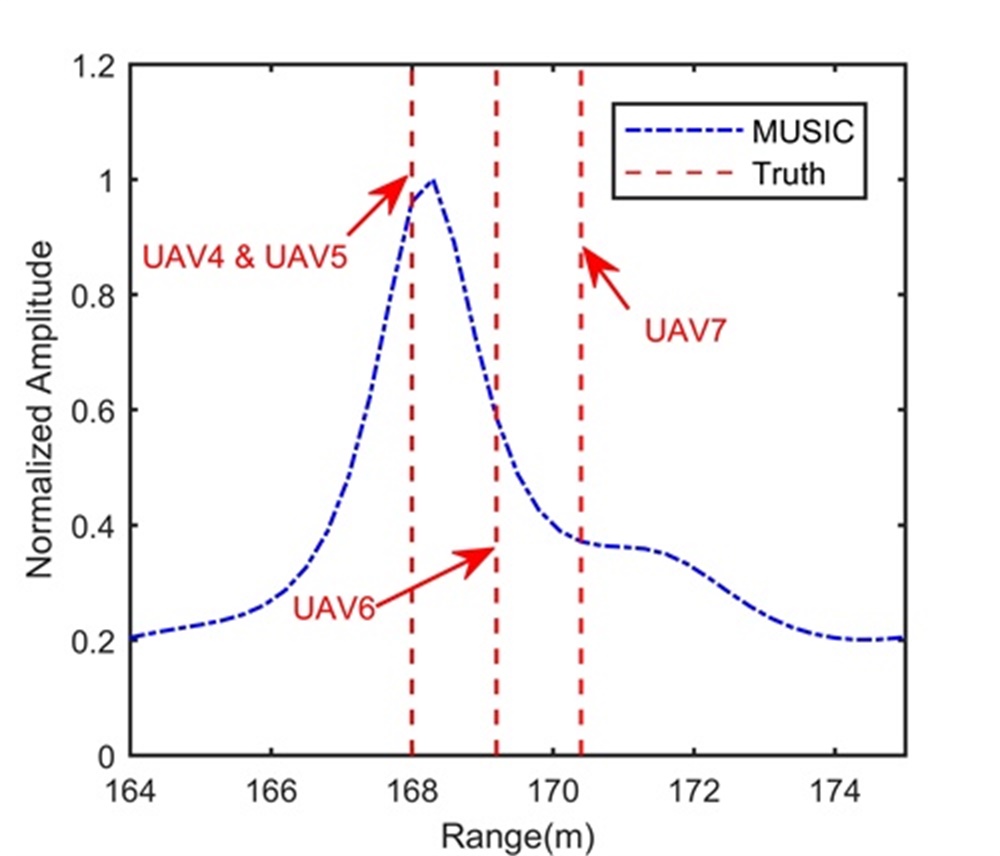}}
		\subfigure[]{
			\label{fig5f}
			\includegraphics[width=55mm]{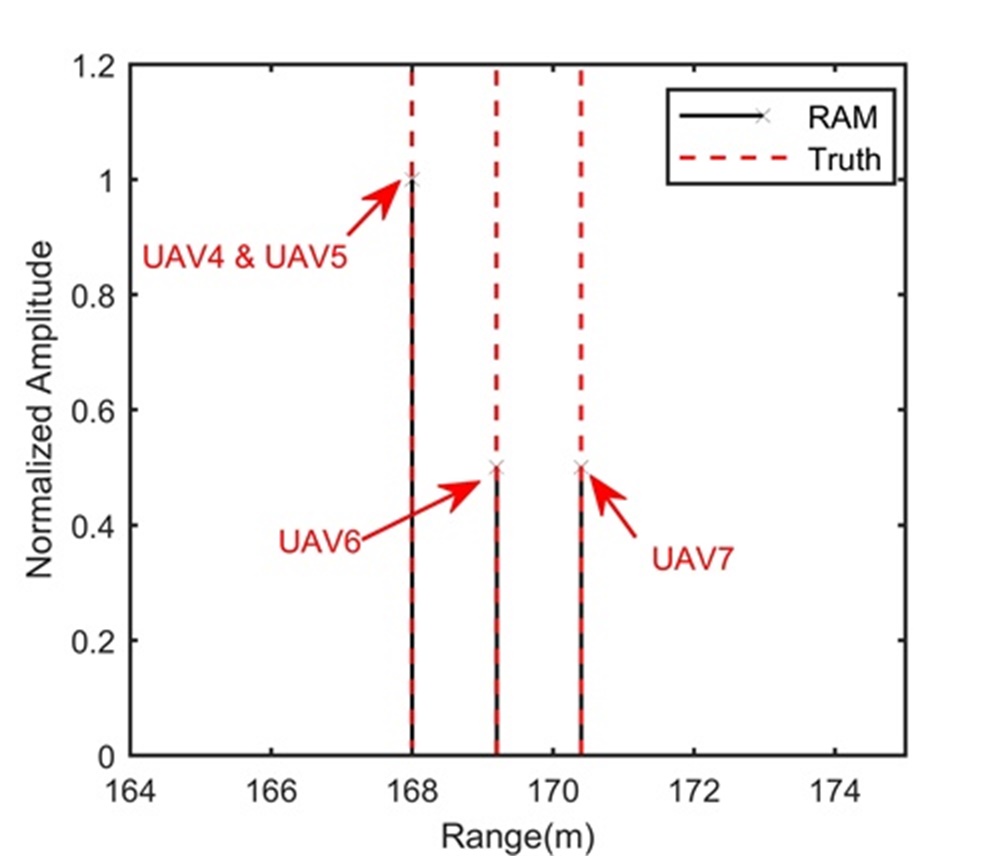}}
		\caption{Radar accurate localization of UAV swarms under noise-free environment. (a) Trajectory in the range-slow time domain after the beamforming operations. (b) Integration result in the range-Doppler domain after \textbf{Step 2}. (c) Integration result in the range-space domain after \textbf{Step 2}. (d) Processing result after \textbf{Step 3}. (e) Processing result of the MUSIC-based method. (f) Processing result of the RAM-based method.}
		\label{fig:Exp2 Result} 
	\end{figure*}

	\textit{\textbf{Experiment 2:}} We consider four UAVs with constant velocities under the noise-free environment. Motion parameters are given in Table \ref{tab:Target para Exp2} and other simulation parameters are same as those in \textit{\textbf{Experiment 1}}. Fig. \ref{fig:Exp2 Result} shows simulation results.

	 Fig. \ref{fig5a} shows the trajectory in the range-slow time domain after the beamforming operation. During the illumination, the RCM happens in Fig. \ref{fig5a} and may influence the LTI. According to radar and target parameters, we know that \textbf{Step 1} cannot separate these four UAVs. Therefore, processing results of \textbf{Step 1} are omitted here. After the processing in \textbf{Step 2}, Figs. \ref{fig5b} and \ref{fig5c} show integration results in the range-Doppler domain and range-space domain, respectively. The longer illumination time let UAV4 separated from UAV5, UAV6 and UAV7 along the Doppler dimension, while we still cannot separate UAV5, UAV6 and UAV7 due to limited range and space resolutions. We extract the signal of the range-Doppler-space bin containing UAV5, UAV6 and UAV7, and apply \textbf{Step 3}. The reconstruction result is shown in Fig. \ref{fig5d}, where UAV5, UAV6 and UAV7 are clearly separated along the range dimension. Therefore, the proposed range super-resolution method completes radar accurate localization of UAV swarms. For comparison, Figs. \ref{fig5e} and \ref{fig5f} show processing results of MUSIC-based method and RAM-based method, respectively. Since the MUSIC-based method has high sidelobe, and is sensitive to the source correlation and loses the Doppler dimension, it performs badly in Fig. \ref{fig5e}. Although the RAM-based method has the low sidelobe and is insensitive to the source correlation, it loses the Doppler dimension also. Therefore, it cannot separate UAV4 from UAV5.
	 
\begin{figure*}[h]
	 	\subfigure[]{
	 		\label{fig6a} 
	 		\includegraphics[width=55mm]{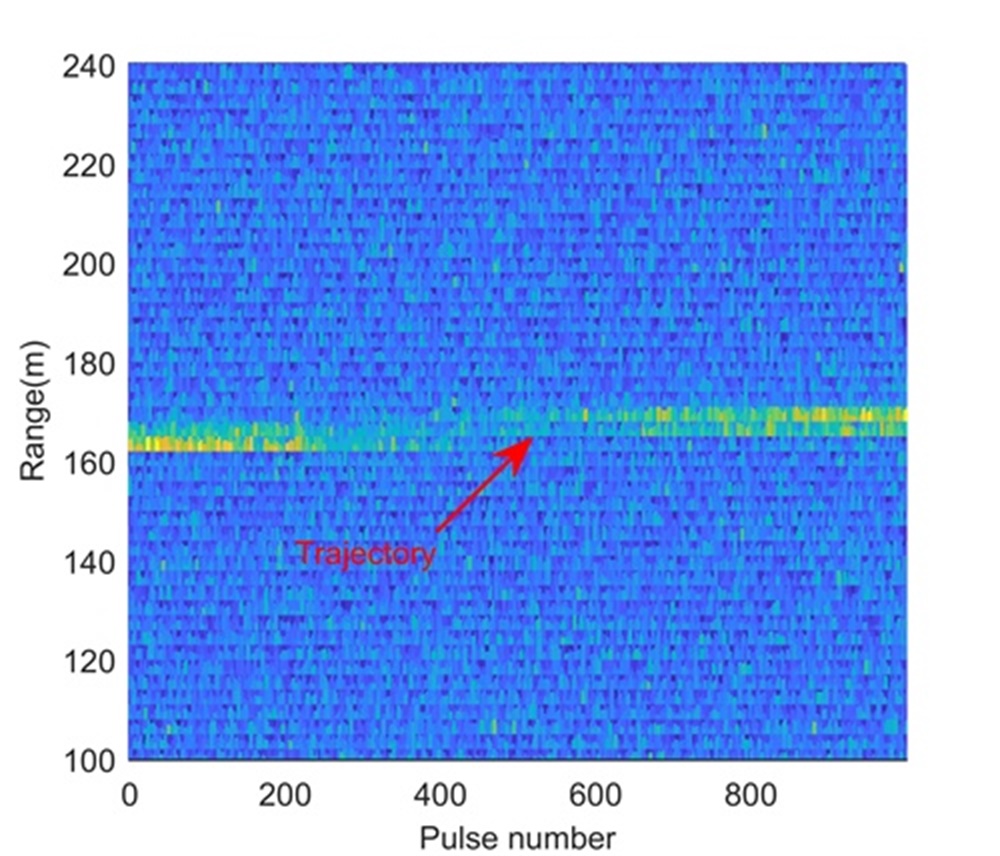}}
	 	\subfigure[]{
	 		\label{fig6b}
	 		\includegraphics[width=55mm]{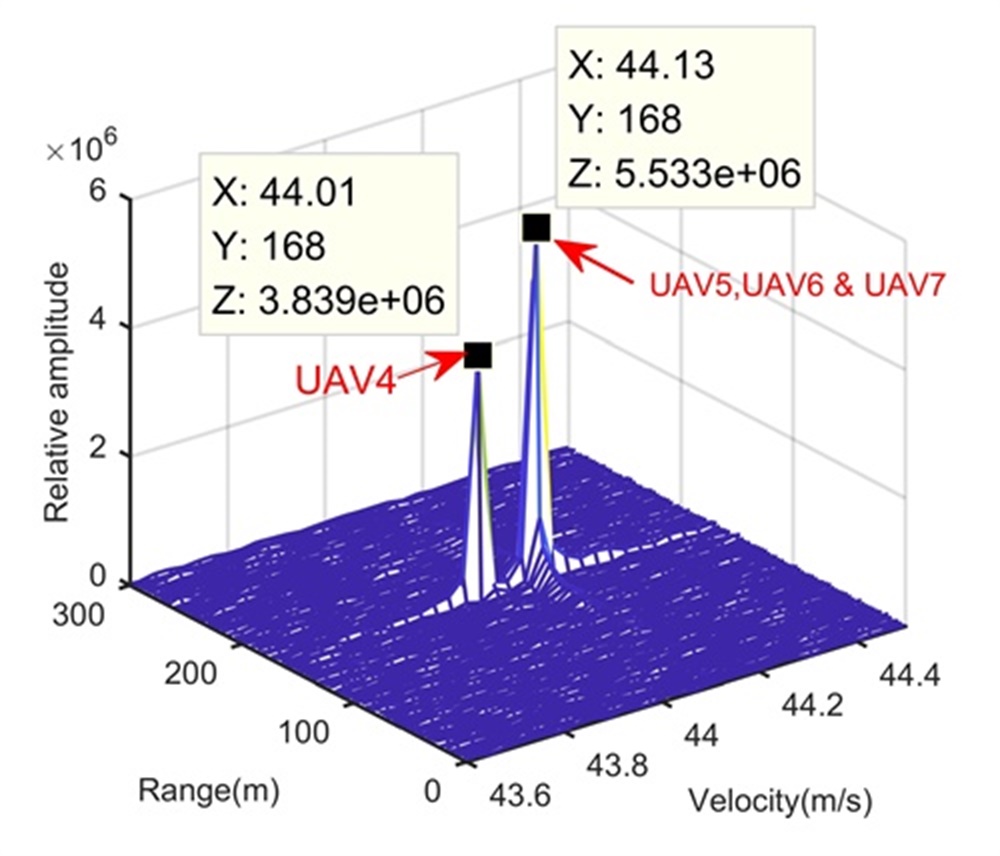}}
	 	\subfigure[]{
	 		\label{fig6c}
	 		\includegraphics[width=55mm]{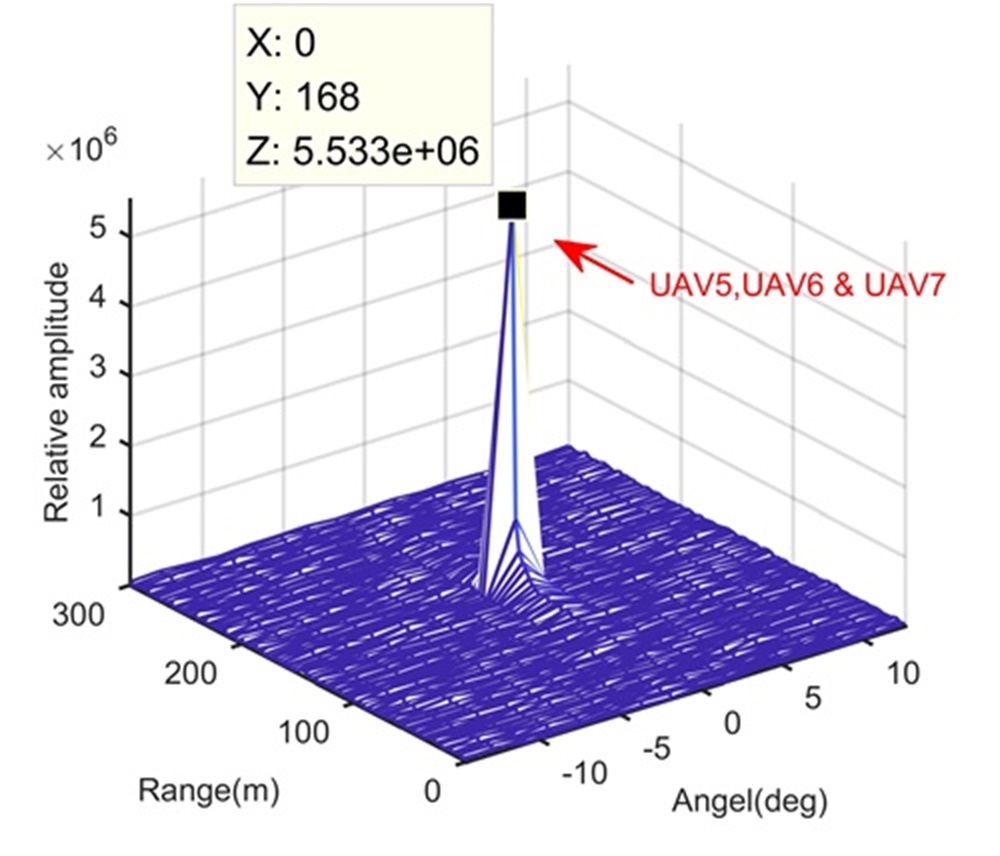}}
	 	
	 	\subfigure[]{
	 		\label{fig6d}
	 		\includegraphics[width=55mm]{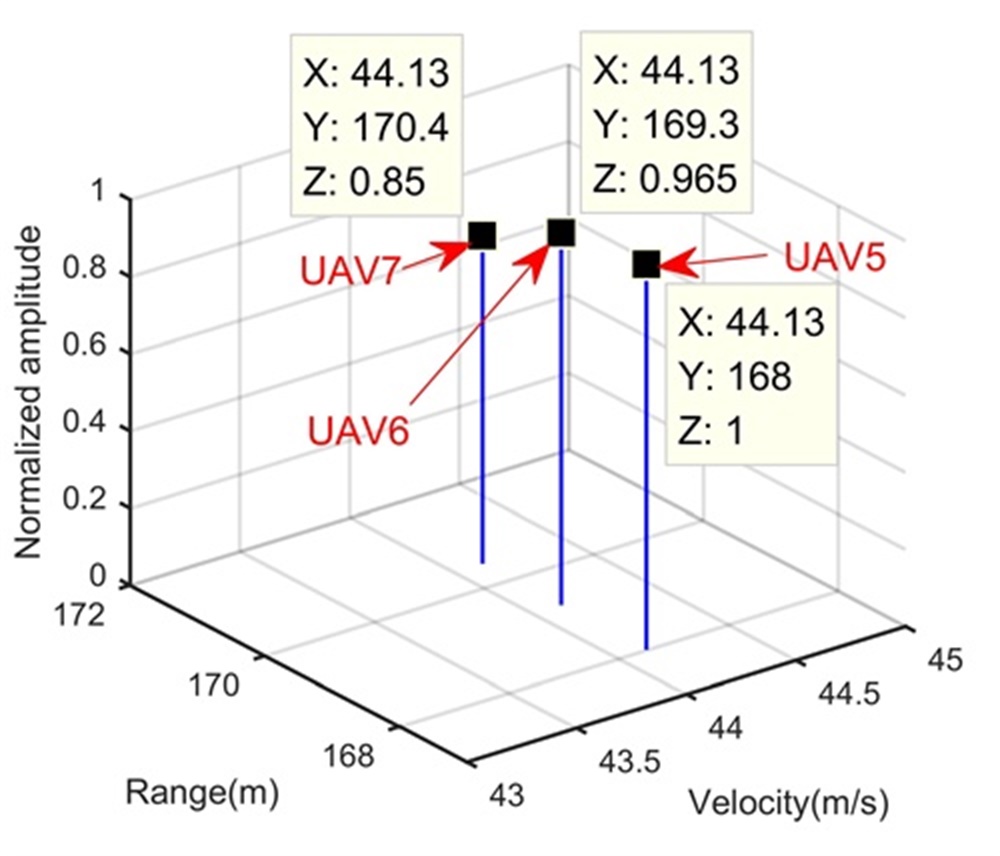}}
	 	\subfigure[]{
	 		\label{fig6e}
	 		\includegraphics[width=55mm]{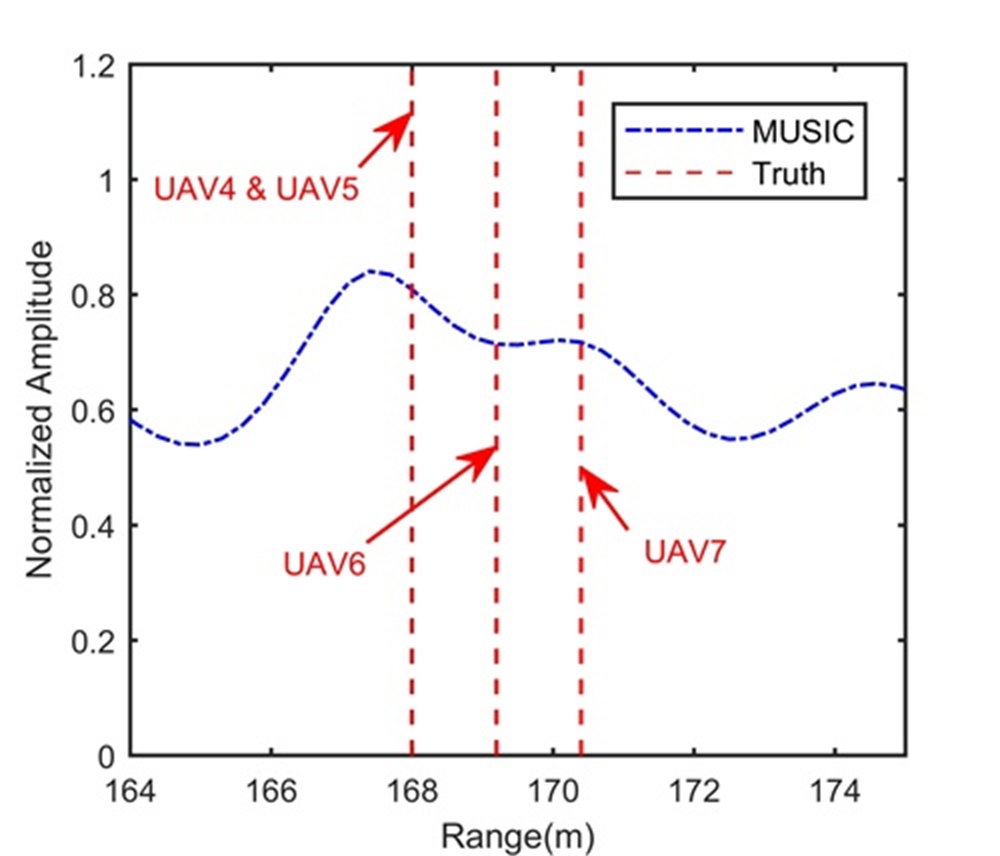}}
	 	\subfigure[]{
	 		\label{fig6f}
	 		\includegraphics[width=55mm]{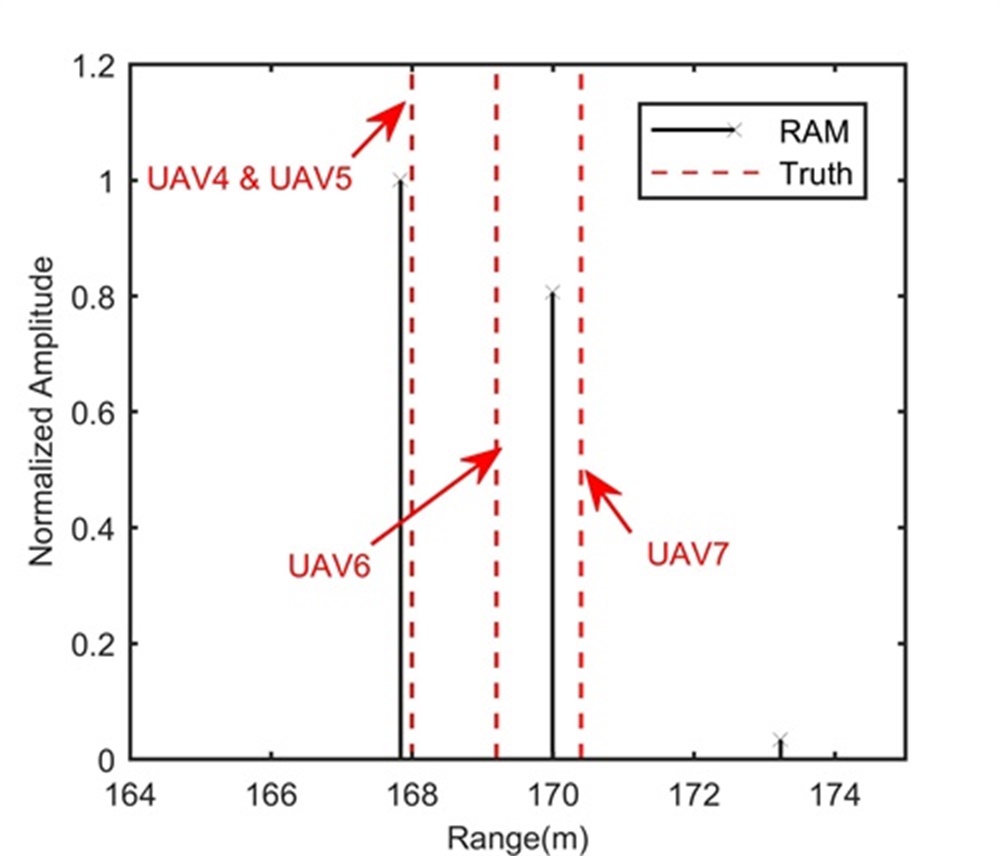}}
	 	\caption{Radar accurate localization of UAV swarms under noise-free environment. (a) Trajectory in the range-slow time domain after the beat and beamforming operations. (b) Integration result in the range-Doppler domain after \textbf{Step 2}. (c) Integration result in the range-space domain after \textbf{Step 2}. (d) Processing result after \textbf{Step 3}. (e) Processing result of the MUSIC-based method. (f) Processing result of the RAM-based method.}
	 	\label{fig:Exp3 Result} 
	 \end{figure*}

	 {\textit{\textbf{Experiment 3:}}} In this experiment, we contaminate the signal with additive complex white Gaussian noise and SNR$ =-13 $dB. Other simulation parameters are the same as those in \textit{\textbf{Experiment 2}}. Fig. \ref{fig:Exp3 Result} shows simulation results.

 	Fig. \ref{fig6a} shows the trajectory in the range-slow time domain after the beamforming operation. After the processing in \textbf{Step 2}, Figs. \ref{fig6b} and \ref{fig6c} show integration results in the range-Doppler domain and range-space domain, respectively. Obviously, compared to Fig. \ref{fig6a}, the LTI greatly enhances the SNR in Figs. \ref{fig6b} and \ref{fig6c}, which benefits the FSRAM in \textbf{Step 3}. Same as processing results in Experiment 2, the long illumination time lets UAV4 separated from UAV5, UAV6 and UAV7 along the Doppler dimension, while UAV5, UAV6 and UAV7 still stay in the same range, Doppler and space cells. We extract the signal of the range-Doppler-space bin containing UAV5, UAV6 and UAV7, and apply \textbf{Step 3}. As expected, UAV5, UAV6 and UAV7 are separated along the range dimension in Fig. \ref{fig6d}. For comparison, Figs. \ref{fig6e} and \ref{fig6f} show processing results of MUSIC-based method and RAM-based method, respectively, which are worse than those in Figs. \ref{fig5e} and \ref{fig5f}. This is because, compared to \textit{\textbf{Experiment 2}}, \textit{\textbf{Experiment 3}} considers the noisy environment. The MUSIC-based method and RAM-based method can only use one modulation period of $ S'_{beat}(n,m',l) $ and are very sensitive to noise. 
 	
 	\subsection{Success rate of adjacent UAVs’ separation}
 	In this subsection, we compare the success rate of adjacent UAVs’ separation. Here, two experiments, i.e., \textit{\textbf{Experiment 4}} and \textit{\textbf{Experiment 5}}, are given. \textit{\textbf{Experiment 4}} is for comparisons among the proposed method, MUSCI-based method and RAM-based method under a fixed SNR, and \textit{\textbf{Experiment 5}} is to illustrate the noise robustness of the proposed method. Note that, since this subsection considers the success rate of adjacent UAVs’ separation and the Doppler information cannot be used by the MUSCI-based method and RAM-based method, we set the same Doppler and space the same for considered UAVs and only their ranges are different. 
 	
\begin{figure*}[h]
 		\centering
 		\subfigure[]{
 			\label{fig7a} 
 			\includegraphics[width=50mm]{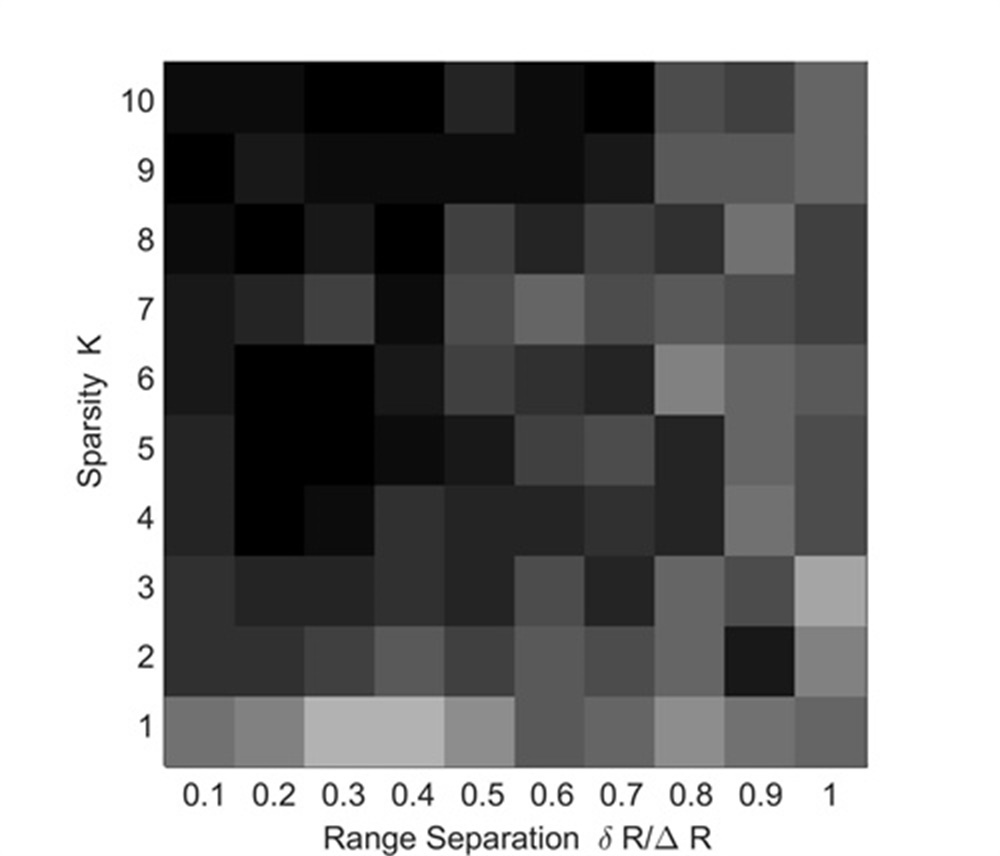}}
 		\subfigure[]{
 			\label{fig7b}
 			\includegraphics[width=50mm]{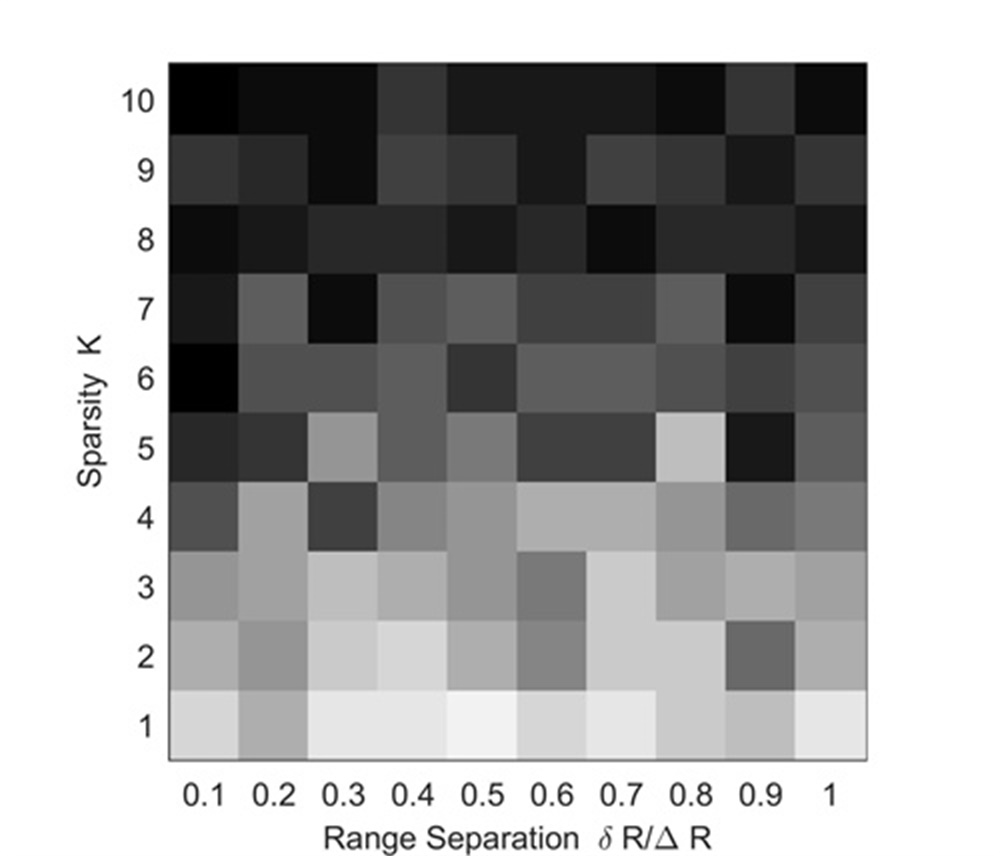}}
 		\subfigure[]{
 			\label{fig7c}
 			\includegraphics[width=50mm]{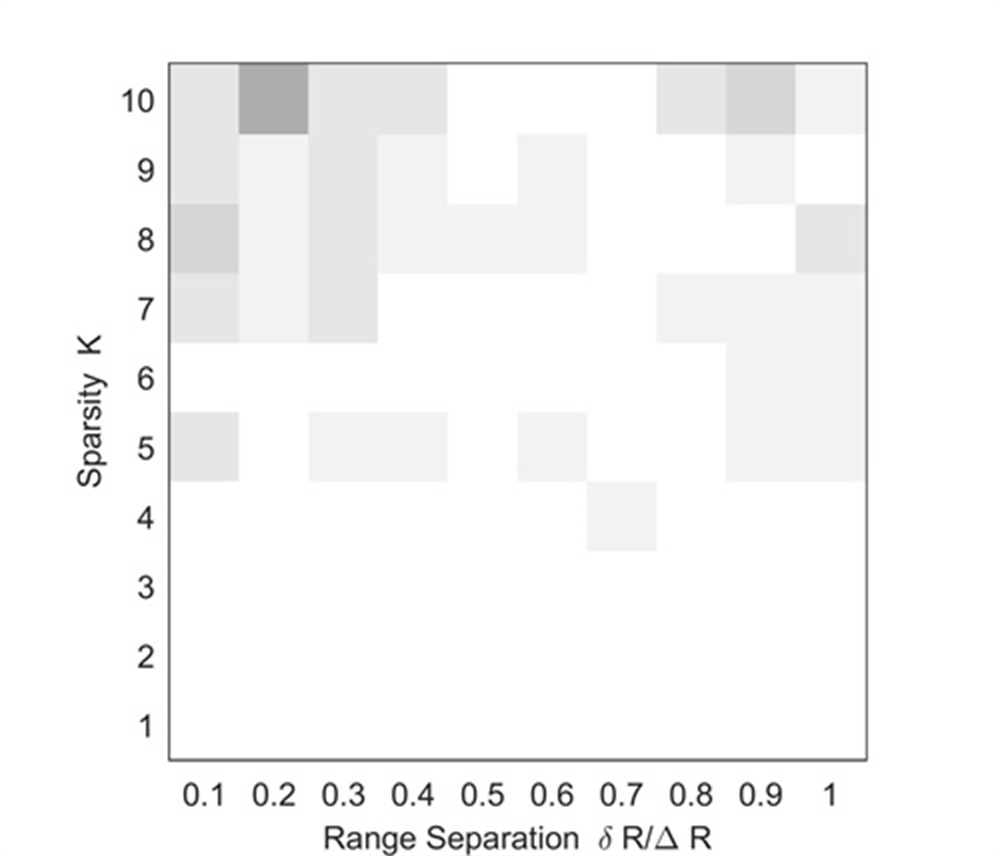}}
 		\caption{Success rate of adjacent UAVs’ separation (a) MUSIC-based method, (b) RAM-based method and (c) proposed method.}
 		\label{fig:Exp4 Result} 
 	\end{figure*}

 	\textit{\textbf{Experiment 4:}} We vary the duo $ (K,\delta R) $  and for each combination we randomly generate ranges such that they are mutually separated by at least $ \delta R /\Delta R $. In this experiment, we only extract 32 samplings along the range dimension for easy comparison. The SNR of $ s_{r}\left(\hat{t}, t_{m}, l\right) $ is 0dB and all UAVs have the same velocity. Other simulation parameters are the same as those in TABLE I. The recovery is called successful if the MSE (Mean Square Error) of the range recovery is less than $ 0.1 \times \Delta R $. For each combination, the success rate is measured over 20 Monte Carlo runs. Fig.\ref{fig:Exp4 Result} shows simulation results.
 	
 	Figs. \ref{fig7a} and \ref{fig7b} show recoveries of the MUSIC-based method and RAM-based method. Since they can only use one modulation period of $ S'_{beat}(n,m',l) $ and are very sensitive to noise, their success rates of adjacent UAVs’ separation are low. Comparing Fig. \ref{fig5f}, Fig. \ref{fig6f} and Fig. \ref{fig7b}, we know that, the RAM-based method may have a high resolution and low sidelobe, while it is very sensitive to the noise. This may greatly influence its realistic applications. The recovery of the proposed method is shown in Fig. \ref{fig7c}, where a high success rate is obtained. This is because, due to the combination of the LTI technique and FSRAM, the proposed method inherits advantages of the noise robustness, low sidelobe and high resolution. 
 	
\begin{figure*}[h]
 		\centering
 		\subfigure[]{
 			\label{fig8a} 
 			\includegraphics[width=40mm]{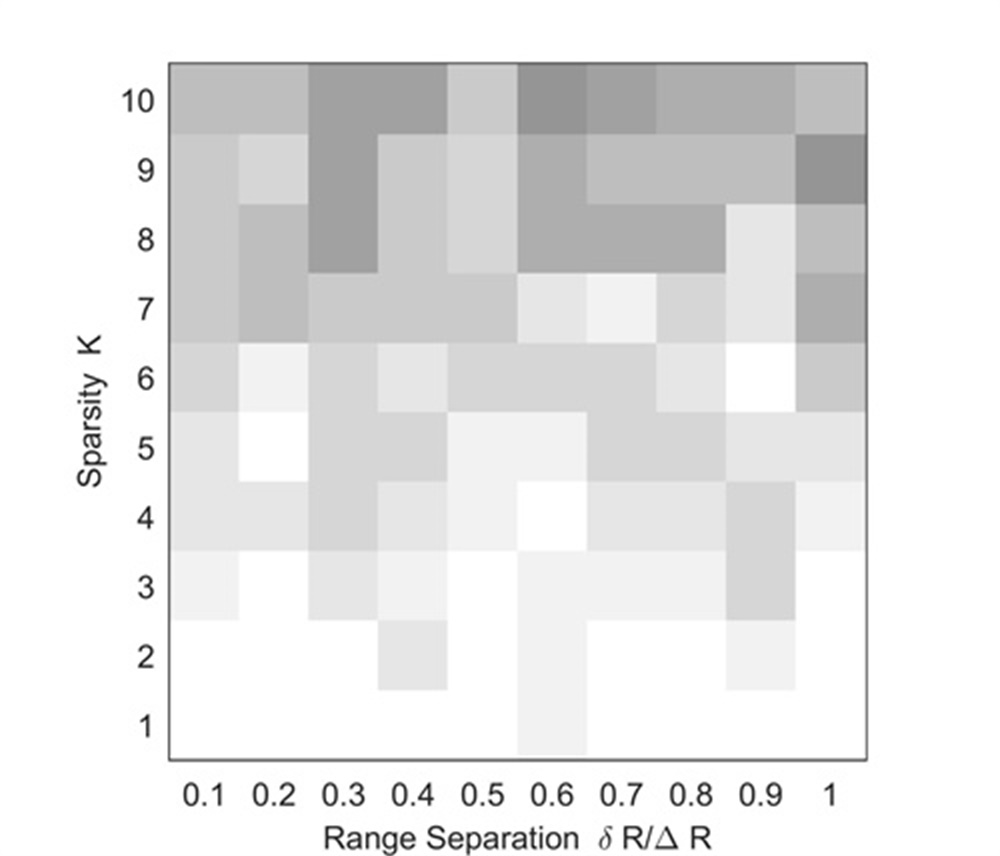}}
 		\subfigure[]{
 			\label{fig8b}
 			\includegraphics[width=40mm]{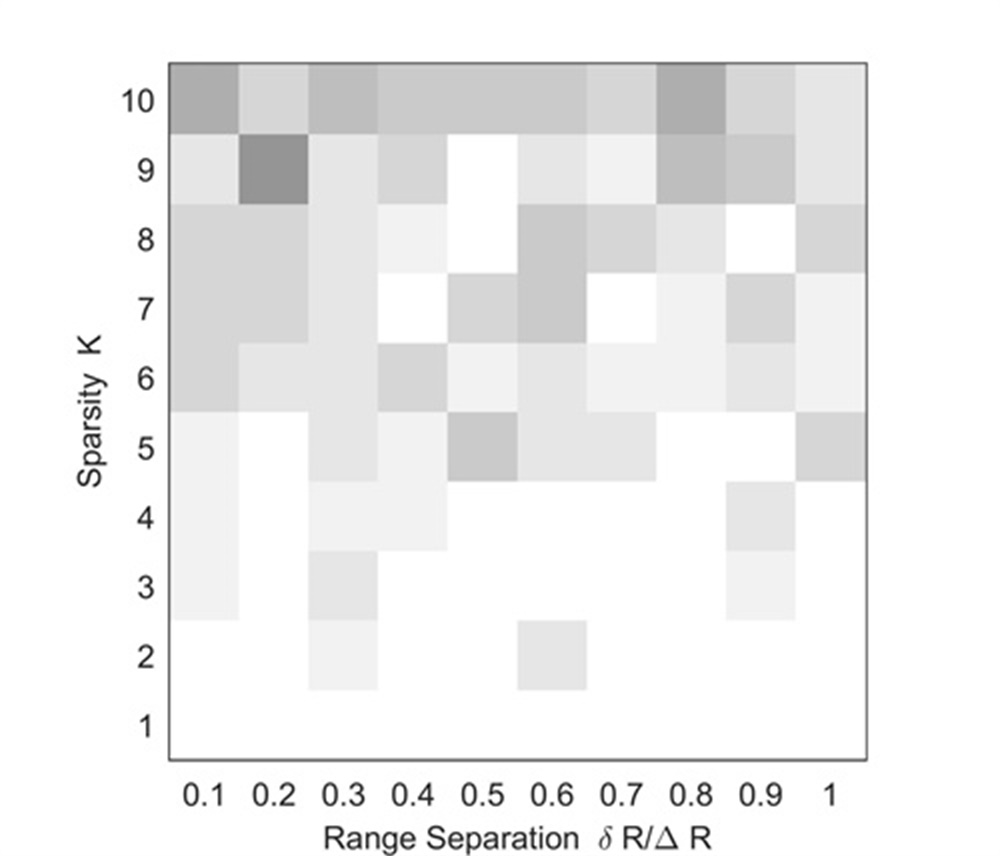}}
 		\subfigure[]{
 			\label{fig8c}
 			\includegraphics[width=40mm]{fig7c.jpg}}
 		\subfigure[]{
 			\label{fig8d}
 			\includegraphics[width=40mm]{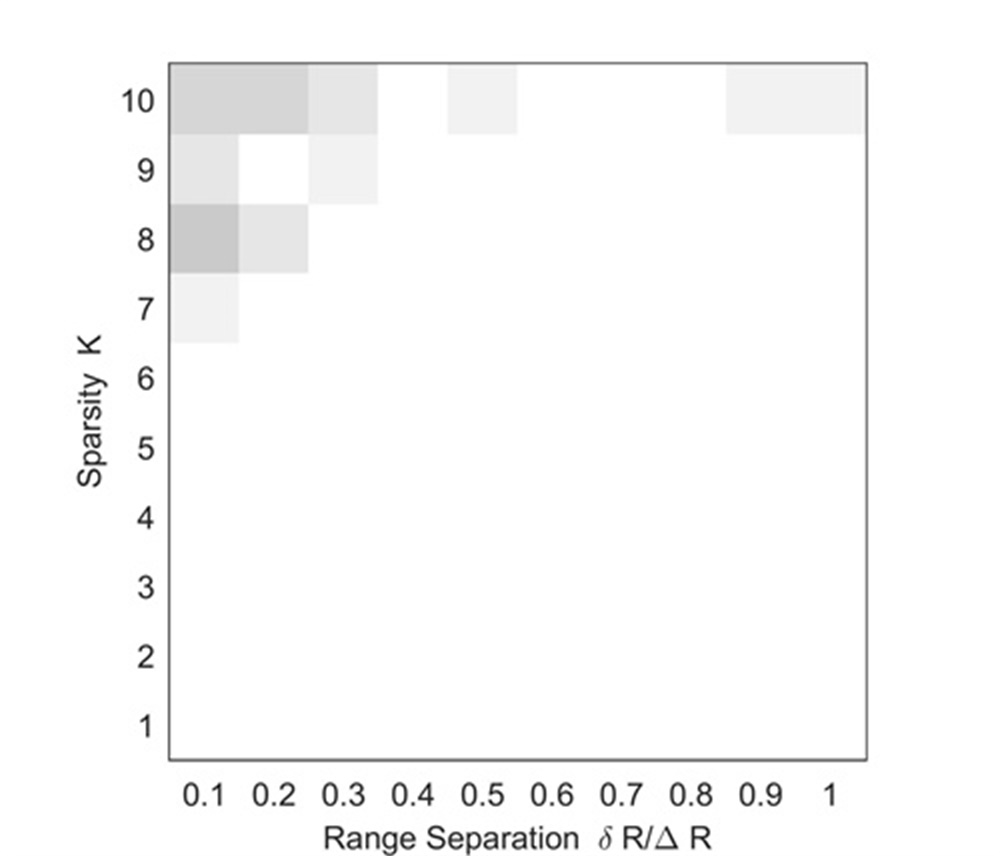}}
 		\caption{Noise robustness illustration of the proposed method (a) Result under the SNR of $ s_r(\hat{t},t_m,l) $  equaling to -20dB. (b) Result under the SNR of $ s_r(\hat{t},t_m,l) $  equaling to -10dB. (c) Result under the SNR of $ s_r(\hat{t},t_m,l) $  equaling to 0dB and (d) Result under the SNR of $ s_r(\hat{t},t_m,l) $  equaling to 10dB.}
 		\label{fig:Exp5 Result} 
 	\end{figure*}

 	\textit{\textbf{Experiment 5:}} Here, we use an experiment to illustrate the noise robustness of the proposed range super-resolution method. Other simulation parameters are the same as those of \textit{\textbf{Experiment 4}}. Figs. \ref{fig8a},\ref{fig8b},\ref{fig8c} and \ref{fig8d} show simulation results under SNRs of $ s_{r}\left(\hat{t}, t_{m}, l\right) $  equaling to -20dB, -10dB, 0dB and 10dB, respectively.
 
 	As can be seen from Fig. \ref{fig:Exp5 Result}, the successful recovery can be obtained easier in the case of a smaller $ K $, a larger $ \delta R $ and a high SNR, which conforms to characteristics of the sparse gridless method. Fig. \ref{fig:Exp5 Result} shows that, when the SNR is larger than 0 dB, the proposed method can complete more than four UAVs’ accurate localizations in one range bin. In this simulation, we only consider 32 samplings along the range dimension for easy comparison. As analyzed in\cite{yang2018sparse,malioutov2005sparse,donoho2009message,zhang2011iterative,zhu2011sparsity_offgrid,fang2016super_offgrid,hu2013fast,tang2013compressed,yang2016exact,candes2014towards,yang2015enhancing}, we know that, if we increase the samplings along the range dimension, the noise robustness of the proposed method will be significantly enhanced. With simulation results shown in Figs. \ref{fig:Exp2 Result}, \ref{fig:Exp3 Result}, \ref{fig:Exp4 Result} and \ref{fig:Exp5 Result}, we have that the proposed method inherits advantages of the LTI technique and sparse gidless method, and are more practical for radar accurate localization of UAV swarms than the LTI technique, MUSIC-based method and RAM-based method.
 	
 	\subsection{Real radar experiment}
 	
\begin{figure*}[h]
 		\centering
 		\subfigure[]{
 			\label{fig9a} 
 			\includegraphics[width=40mm]{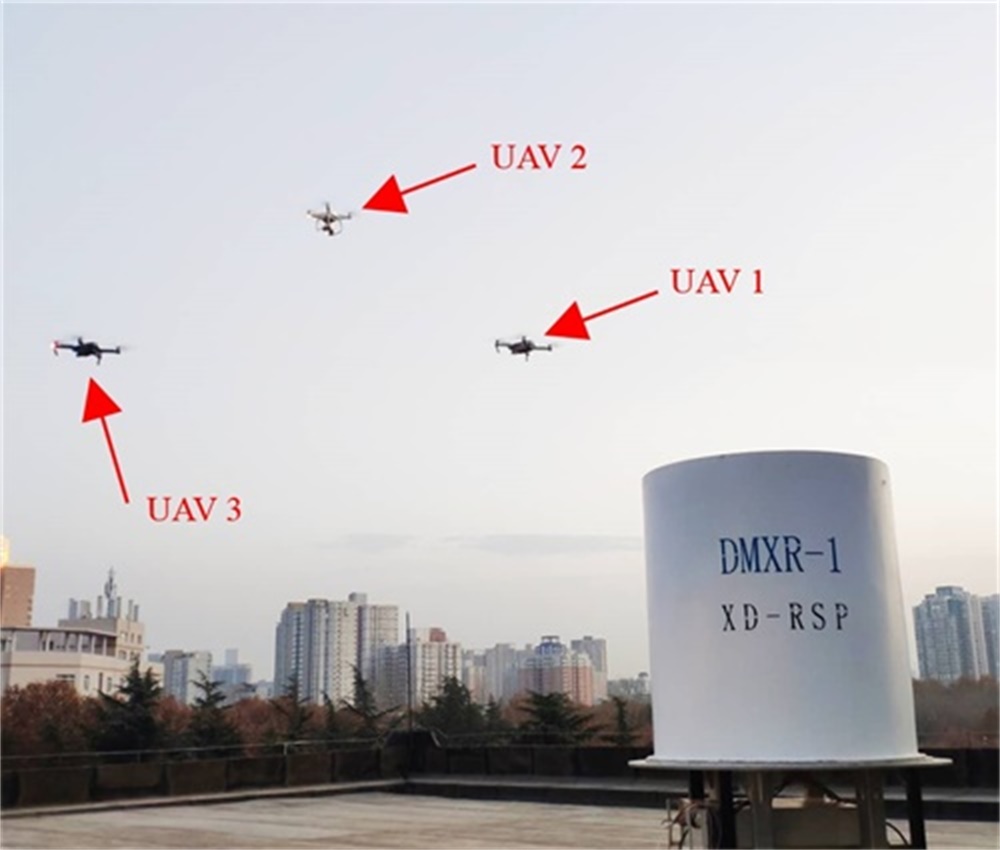}}
 		\subfigure[]{
 			\label{fig9b}
 			\includegraphics[width=44mm]{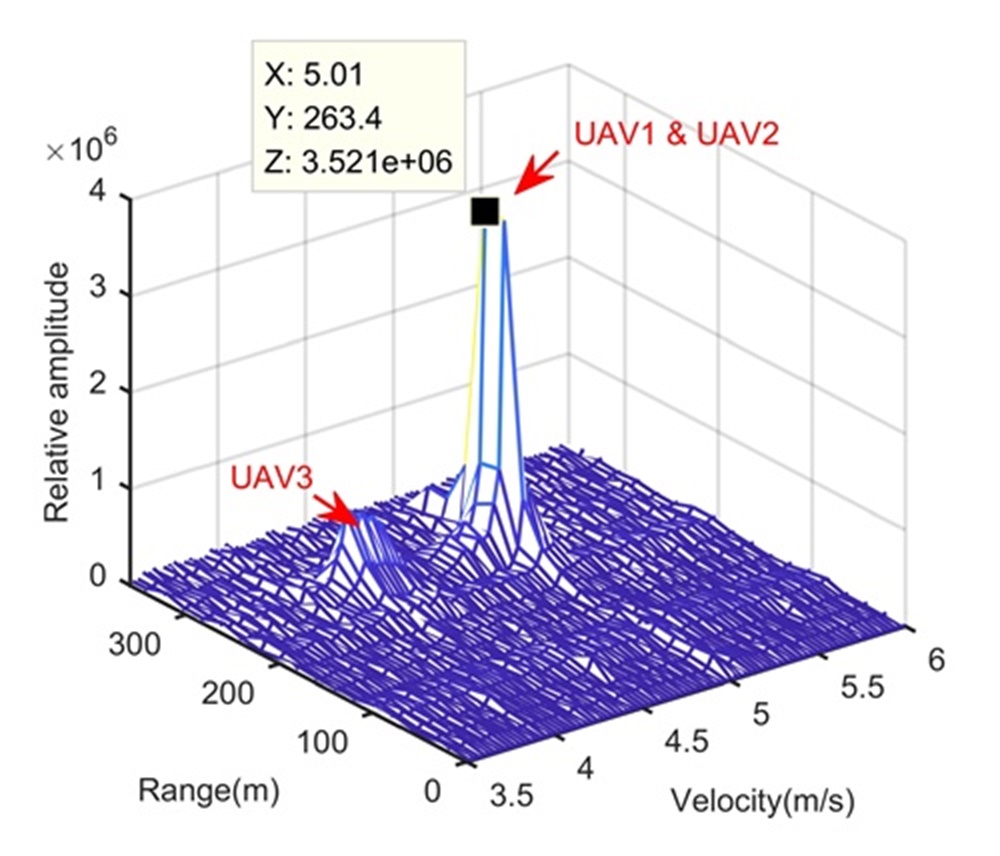}}	
 		\subfigure[]{
 			\label{fig9c}
 			\includegraphics[width=44mm]{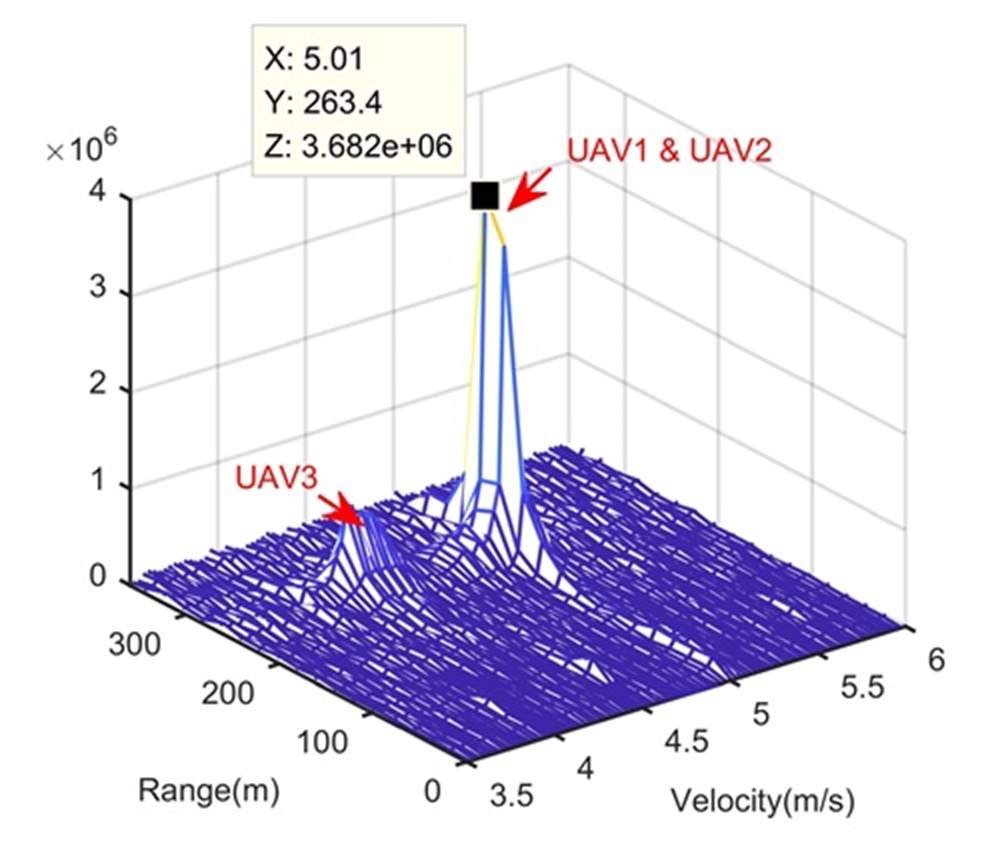}}
 		\subfigure[]{
 			\label{fig9d}
 			\includegraphics[width=44mm]{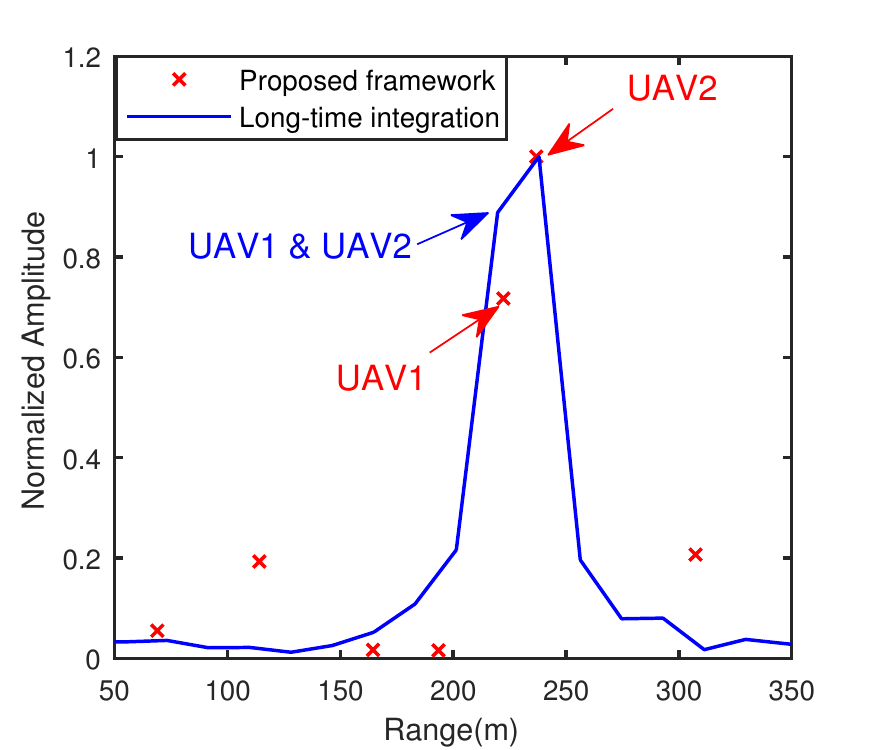}}
 		\caption{Processing results of the real radar data. (a) Scenario of the experiment. (b) Processing result of the tradition moving target detection method. (c) Integration result in the range-Doppler domain after \textbf{ Step 2}. (d) Processing result after \textbf{ Step 3}.}
 		\label{fig:realdata Result} 
 	\end{figure*}

 	In this section, the real data is used to validate the practicability of proposed range super-resolution method. The real data was collected by an X-band radar and three UAVs shown in Fig. \ref{fig9a}. Parameters of this X-band radar are listed in Table \ref{tab:Radar para Real Data}. In this experiment, we let UAV1, UAV2 and UAV3 in the same range cell, UAV1 and UAV2 with the same radial velocity, and UAV3 with a lower radial velocity. Since we have known where the UAV swarms is, we skip \textbf{Step 1} of the proposed method. Fig. \ref{fig:realdata Result} gives the processing results.
 	\begin{table}[h]
 	\centering
 	\caption{Radar Parameters}
 	\begin{tabular}{cc}
 		\toprule
 		Parameter  &Value  \\  	
 		\midrule  
 		Carrier frequency band  &X band  \\  	
 		BandWidth  & 10 MHz   \\
 		Chirp duration    & 500  $\mu s$   \\
 		Sampling frequency      & 10 MHz\\
 		Number of antenna elements    &8     \\
 		Range resolution    & 15  m   \\
 		\bottomrule
 	\end{tabular}\label{tab:Radar para Real Data}
 	\end{table}
 	
 	Figs. \ref{fig9b} and \ref{fig9c} show processing results of the traditional target detection method and LTI method in \textbf{Step 2}, respectively. These processing results can illustrate that i) during the illumination time, the RCM happens to three UAVs; ii) the SNR of the echo is low and the LTI method in \textbf{Step 2} significantly enhance the SNR; iii) the range resolution is low and cannot separate UAV1 and UAV2. According to \textbf{Step 3} of the proposed method, we extract the data and use the FSRAM to separate UAV1 and UAV2. The processing result is shown in Fig. \ref{fig9d}, where the extracted data along the range dimension is also shown for comparison. Obviously, the proposed method succeeds to separate UAV1 and UAV2. Simulation results in Fig. \ref{fig:realdata Result} can demonstrate the practicability of proposed range super-resolution method in realistic applications.

	\section{Conclusion and Future Work}\label{Conclusion}
	In this paper, we study radar accurate localization of UAV swarms based on a range super-resolution method. First, for characteristics of UAV swarms, we construct a super-resolution framework by using advantages of the long-time integration (LTI) technique and gridless sparse method. Thereafter, based on this framework, a range super-resolution method is proposed to realize the radar accurate localization of UAV swarms. Mathematical analyses and numerical simulations are performed to demonstrate superiorities of the proposed range super-resolution method. Compared to the KT-based LTI method, MUSIC-based method and RAM-based method, the proposed method is more suitable for radar accurate localization of UAV swarms. The real experiment with X-band radar is also conducted to verify the effectiveness of the range super-resolution method.

	The accurate defense of the UAV swarms is more and more important right now. In this paper, we only consider the UAV swarms with constant velocities. However, in realistic applications, the UAV swarms with accelerations may happen. Based on the proposed framework, more advanced methods should be studied to deal with this situation more effectively. In addition, the efficient implementation of the sparse method is critical. In future, we will study  more computationally efficient sparse method.
\subsubsection*{Acknowledgments.} The authors would like to thank Dr. Zai Yang of Xi'an Jiaotong University, China, for helpful discussions on the gridless sparse method.

	\bibliographystyle{IEEEtran}
	\bibliography{References}

\begin{thebibliography}{10}
\providecommand{\url}[1]{#1}
\csname url@samestyle\endcsname
\providecommand{\newblock}{\relax}
\providecommand{\bibinfo}[2]{#2}
\providecommand{\BIBentrySTDinterwordspacing}{\spaceskip=0pt\relax}
\providecommand{\BIBentryALTinterwordstretchfactor}{4}
\providecommand{\BIBentryALTinterwordspacing}{\spaceskip=\fontdimen2\font plus
\BIBentryALTinterwordstretchfactor\fontdimen3\font minus
  \fontdimen4\font\relax}
\providecommand{\BIBforeignlanguage}[2]{{%
\expandafter\ifx\csname l@#1\endcsname\relax
\typeout{** WARNING: IEEEtran.bst: No hyphenation pattern has been}%
\typeout{** loaded for the language `#1'. Using the pattern for}%
\typeout{** the default language instead.}%
\else
\language=\csname l@#1\endcsname
\fi
#2}}
\providecommand{\BIBdecl}{\relax}
\BIBdecl

\bibitem{wei2013operation}
Y.~Wei, M.~B. Blake, and G.~R. Madey, ``{An operation-time simulation framework
  for UAV swarm configuration and mission planning},'' \emph{Procedia Computer
  Science}, vol.~18, pp. 1949--1958, 2013.

\bibitem{daponte2015metrology}
P.~Daponte, L.~De~Vito, G.~Mazzilli, F.~Picariello, S.~Rapuano, and M.~Riccio,
  ``{Metrology for drone and drone for metrology: Measurement systems on small
  civilian drones},'' in \emph{2015 IEEE Metrology for Aerospace
  (MetroAeroSpace)}.\hskip 1em plus 0.5em minus 0.4em\relax IEEE, 2015, pp.
  306--311.

\bibitem{bacco2016satellites}
M.~Bacco, L.~Caviglione, and A.~Gotta, ``Satellites, uavs, vehicles and sensors
  for an integrated delay tolerant ad hoc network,'' in \emph{International
  Conference on Personal Satellite Services}.\hskip 1em plus 0.5em minus
  0.4em\relax Springer, 2016, pp. 114--122.

\bibitem{albani2017field}
D.~Albani, D.~Nardi, and V.~Trianni, ``{Field coverage and weed mapping by UAV
  swarms},'' in \emph{2017 IEEE/RSJ International Conference on Intelligent
  Robots and Systems (IROS)}.\hskip 1em plus 0.5em minus 0.4em\relax IEEE,
  2017, pp. 4319--4325.

\bibitem{wu2018modeling}
H.~Wu, H.~Li, R.~Xiao, and J.~Liu, ``{Modeling and simulation of dynamic ant
  colony’s labor division for task allocation of UAV swarm},'' \emph{Physica
  A: Statistical Mechanics and its Applications}, vol. 491, pp. 127--141, 2018.

\bibitem{grossi2016new}
E.~Grossi, M.~Lops, and L.~Venturino, ``A new look at the radar detection
  problem,'' \emph{IEEE Transactions on Signal Processing}, vol.~64, no.~22,
  pp. 5835--5847, 2016.

\bibitem{xu2017focus}
J.~Xu, Y.-N. Peng, X.-G. Xia, and A.~Farina, ``{Focus-before-detection radar
  signal processing: Part I—Challenges and methods},'' \emph{IEEE Aerospace
  and Electronic Systems Magazine}, vol.~32, no.~9, pp. 48--59, 2017.

\bibitem{carlson1994search}
B.~Carlson, E.~Evans, and S.~Wilson, ``{Search radar detection and track with
  the Hough transform. I. system concept},'' \emph{IEEE transactions on
  aerospace and electronic systems}, vol.~30, no.~1, pp. 102--108, 1994.

\bibitem{carretero2009application}
J.~Carretero-Moya, J.~Gismero-Menoyo, A.~Asensio-Lopez, and A.~Blanco-del
  Campo, ``Application of the radon transform to detect small-targets in sea
  clutter,'' \emph{IET radar, sonar \& navigation}, vol.~3, no.~2, pp.
  155--166, 2009.

\bibitem{tandra2008snr}
R.~Tandra and A.~Sahai, ``{SNR walls for signal detection},'' \emph{IEEE
  Journal of selected topics in Signal Processing}, vol.~2, no.~1, pp. 4--17,
  2008.

\bibitem{abatzoglou1998range}
T.~J. Abatzoglou and G.~O. Gheen, ``{Range, radial velocity, and acceleration
  MLE using radar LFM pulse train},'' \emph{IEEE Transactions on Aerospace and
  Electronic Systems}, vol.~34, no.~4, pp. 1070--1083, 1998.

\bibitem{yu2012radon}
J.~Yu, J.~Xu, Y.-N. Peng, and X.-G. Xia, ``{Radon-Fourier transform for radar
  target detection (III): optimality and fast implementations},'' \emph{IEEE
  Transactions on Aerospace and Electronic Systems}, vol.~48, no.~2, pp.
  991--1004, 2012.

\bibitem{zhu2007keystone}
D.~Zhu, Y.~Li, and Z.~Zhu, ``{A keystone transform without interpolation for
  SAR ground moving-target imaging},'' \emph{IEEE Geoscience and Remote Sensing
  Letters}, vol.~4, no.~1, pp. 18--22, 2007.

\bibitem{zheng2015radar}
J.~Zheng, T.~Su, H.~Liu, G.~Liao, Z.~Liu, and Q.~H. Liu, ``Radar high-speed
  target detection based on the frequency-domain deramp-keystone transform,''
  \emph{IEEE Journal of Selected Topics in Applied Earth Observations and
  Remote Sensing}, vol.~9, no.~1, pp. 285--294, 2015.

\bibitem{zheng2017parameterized}
J.~Zheng, H.~Liu, and Q.~H. Liu, ``{Parameterized centroid frequency-chirp rate
  distribution for LFM signal analysis and mechanisms of constant delay
  introduction},'' \emph{IEEE Transactions on Signal Processing}, vol.~65,
  no.~24, pp. 6435--6447, 2017.

\bibitem{schmidt1986multiple}
R.~Schmidt, ``Multiple emitter location and signal parameter estimation,''
  \emph{IEEE transactions on antennas and propagation}, vol.~34, no.~3, pp.
  276--280, 1986.

\bibitem{paulraj1986subspace}
A.~Paulraj, R.~Roy, and T.~Kailath, ``A subspace rotation approach to signal
  parameter estimation,'' \emph{Proceedings of the IEEE}, vol.~74, no.~7, pp.
  1044--1046, 1986.

\bibitem{yang2018sparse}
Z.~Yang, J.~Li, P.~Stoica, and L.~Xie, ``Sparse methods for
  direction-of-arrival estimation,'' in \emph{Academic Press Library in Signal
  Processing, Volume 7}.\hskip 1em plus 0.5em minus 0.4em\relax Elsevier, 2018,
  pp. 509--581.

\bibitem{malioutov2005sparse}
D.~Malioutov, M.~Cetin, and A.~S. Willsky, ``A sparse signal reconstruction
  perspective for source localization with sensor arrays,'' \emph{IEEE
  transactions on signal processing}, vol.~53, no.~8, pp. 3010--3022, 2005.

\bibitem{donoho2009message}
D.~L. Donoho, A.~Maleki, and A.~Montanari, ``Message-passing algorithms for
  compressed sensing,'' \emph{Proceedings of the National Academy of Sciences},
  vol. 106, no.~45, pp. 18\,914--18\,919, 2009.

\bibitem{zhang2011iterative}
Z.~Zhang and B.~D. Rao, ``Iterative reweighted algorithms for sparse signal
  recovery with temporally correlated source vectors,'' in \emph{2011 IEEE
  International Conference on Acoustics, Speech and Signal Processing
  (ICASSP)}.\hskip 1em plus 0.5em minus 0.4em\relax IEEE, 2011, pp. 3932--3935.

\bibitem{zhu2011sparsity_offgrid}
H.~Zhu, G.~Leus, and G.~B. Giannakis, ``Sparsity-cognizant total least-squares
  for perturbed compressive sampling,'' \emph{IEEE Transactions on Signal
  Processing}, vol.~59, no.~5, pp. 2002--2016, 2011.

\bibitem{fang2016super_offgrid}
J.~Fang, F.~Wang, Y.~Shen, H.~Li, and R.~S. Blum, ``{Super-resolution
  compressed sensing for line spectral estimation: An iterative reweighted
  approach},'' \emph{IEEE Transactions on Signal Processing}, vol.~64, no.~18,
  pp. 4649--4662, 2016.

\bibitem{hu2013fast}
L.~Hu, J.~Zhou, Z.~Shi, and Q.~Fu, ``A fast and accurate reconstruction
  algorithm for compressed sensing of complex sinusoids,'' \emph{IEEE
  Transactions on Signal Processing}, vol.~61, no.~22, pp. 5744--5754, 2013.

\bibitem{tang2013compressed}
G.~Tang, B.~N. Bhaskar, P.~Shah, and B.~Recht, ``Compressed sensing off the
  grid,'' \emph{IEEE transactions on information theory}, vol.~59, no.~11, pp.
  7465--7490, 2013.

\bibitem{yang2016exact}
Z.~Yang and L.~Xie, ``Exact joint sparse frequency recovery via optimization
  methods,'' \emph{IEEE Transactions on Signal Processing}, vol.~64, no.~19,
  pp. 5145--5157, 2016.

\bibitem{candes2014towards}
E.~J. Cand{\`e}s and C.~Fernandez-Granda, ``Towards a mathematical theory of
  super-resolution,'' \emph{Communications on pure and applied Mathematics},
  vol.~67, no.~6, pp. 906--956, 2014.

\bibitem{yang2015enhancing}
Z.~Yang and L.~Xie, ``Enhancing sparsity and resolution via reweighted atomic
  norm minimization,'' \emph{IEEE Transactions on Signal Processing}, vol.~64,
  no.~4, pp. 995--1006, 2015.

\bibitem{mishra2015spectral}
K.~V. Mishra, M.~Cho, A.~Kruger, and W.~Xu, ``Spectral super-resolution with
  prior knowledge,'' \emph{IEEE transactions on signal processing}, vol.~63,
  no.~20, pp. 5342--5357, 2015.

\bibitem{yang2018frequency}
Z.~Yang and L.~Xie, ``{Frequency-selective Vandermonde decomposition of
  Toeplitz matrices with applications},'' \emph{Signal Processing}, vol. 142,
  pp. 157--167, 2018.

\bibitem{zhang2018multitarget}
D.~Zhang, Y.~He, X.~Gong, Y.~Hu, Y.~Chen, and B.~Zeng, ``{Multitarget AOA
  estimation using wideband LFMCW signal and two receiver antennas},''
  \emph{IEEE Transactions on Vehicular Technology}, vol.~67, no.~8, pp.
  7101--7112, 2018.

\bibitem{brooker2005understanding}
G.~M. Brooker, ``Understanding millimetre wave fmcw radars,'' in \emph{1st
  international Conference on Sensing Technology}, 2005, pp. 152--157.

\bibitem{pieraccini2017rcs}
M.~Pieraccini, L.~Miccinesi, and N.~Rojhani, ``{RCS measurements and ISAR
  images of small UAVs},'' \emph{IEEE Aerospace and Electronic Systems
  Magazine}, vol.~32, no.~9, pp. 28--32, 2017.

\bibitem{toh1999sdpt3}
K.-C. Toh, M.~J. Todd, and R.~H. T{\"u}t{\"u}nc{\"u}, ``{SDPT3—a MATLAB
  software package for semidefinite programming, version 1.3},''
  \emph{Optimization methods and software}, vol.~11, no. 1-4, pp. 545--581,
  1999.

\bibitem{boyd2011distributed}
S.~Boyd, N.~Parikh, E.~Chu, B.~Peleato, J.~Eckstein \emph{et~al.},
  ``Distributed optimization and statistical learning via the alternating
  direction method of multipliers,'' \emph{Foundations and
  Trends{\textregistered} in Machine learning}, vol.~3, no.~1, pp. 1--122,
  2011.

\end{thebibliography}
\end{document}